\documentclass[twocolumn]{autart}

\usepackage{amsmath,amssymb,amsfonts}
\usepackage{algorithm}
\usepackage{algpseudocode}
\usepackage{graphicx}
\usepackage{textcomp}
\usepackage{url}
\usepackage{float}

\usepackage{enumitem}
\usepackage[T1]{fontenc}
\usepackage[utf8]{inputenc}
\usepackage[titletoc]{appendix}
\usepackage[authoryear]{natbib}
\usepackage[hidelinks]{hyperref}

\DeclareMathAlphabet{\mathcal}{OMS}{cmsy}{m}{n}

\def\mP{\mathbb{P}}
\def\mE{\mathbb{E}}

\def\as{\text{a.s.}}

\DeclareMathOperator{\diag}{diag}
\DeclareMathOperator{\col}{col}
\DeclareMathOperator{\trace}{trace}

\DeclareMathOperator{\rank}{rank}

\newcommand{\Absl}[1]{\lVert #1 \rVert}

\newcommand{\Abs}[1]{\left\| #1\right\|}

\newtheorem{lemx}{Lemma}[section]

\begin{document}

\begin{frontmatter}
\runtitle{Privacy-Preserving Cram{\'e}r-Rao Lower Bound}

\title{Privacy-Preserving Cram{\'e}r-Rao Lower Bound \thanksref{footnoteinfo}}

\thanks[footnoteinfo]{This work was supported by the National Natural Science Foundation of China under Grants 62433020 and T2293772. Corresponding author: Ji-Feng Zhang.}

\author[padova]{Jieming Ke}\ead{kejieming@amss.ac.cn},
\author[ustb,keylab]{Jimin Wang}\ead{jimwang@ustb.edu.cn},
\author[zyut,amss]{Ji-Feng Zhang}\ead{jif@iss.ac.cn}

%\address[ucas]{School of Mathematical Sciences, University of Chinese Academy of Sciences, Beijing 100049, China}
\address[padova]{Department of Information Engineering, University of Padova, Padova 35131, Italy}
\address[ustb]{School of Automation and Electrical Engineering, University of Science and Technology Beijing, Beijing 100083, China}
\address[keylab]{Key Laboratory of Knowledge Automation for Industrial Processes, Ministry of Education, Beijing 100083, China}
\address[amss]{State Key Laboratory of Mathematical Sciences, Academy of Mathematics and Systems Science, Chinese Academy of Sciences, Beijing 100190, China}
\address[zyut]{School of Automation and Electrical Engineering, Zhongyuan University of Technology, Zhengzhou 450007, Henan Province, China}


\begin{abstract}
This paper establishes the privacy-preserving Cram{\'e}r-Rao lower bound (CRLB) theory, characterizing the fundamental limit of identification accuracy under privacy constraint for general stochastic obfuscation mechanisms. 
An identifiability criterion under privacy constraint is derived by using Fisher information matrix as the privacy metric. 
In the identifiable case, a precise privacy-preserving CRLB is established with an explicit expression, which quantifies the privacy cost without unspecified constant factors. 
Considering computational efficiency, recursive formulas are developed to compute the privacy-preserving CRLB for multi-measurement systems, reducing the computational burden caused by direct high-dimensional matrix inversion. 
To demonstrate the tightness of the lower bound, a Gaussian-mechanism-based privacy-preserving recursive least-squares algorithm is shown to strictly attain the bound under Gaussian measurement noises, and a maximum-likelihood-based privacy-preserving identification algorithm is proposed to attain the bound in the sense of convergence rates under non-Gaussian measurement noises.
%For applicability, the proposed theory can be extended to dynamic model state estimation, distributed estimation, and average consensus.
Experimental results are provided to demonstrate the privacy-preserving CRLB and show the effectiveness of the proposed algorithms.
\vspace{-0.5em}
\end{abstract}

\end{frontmatter}

\section{Introduction}

\subsection{Background and Motivations}

The practical challenge of utilizing sensitive data while preserving privacy motivates the critical research problem of privacy-preserving system identification. For example, \citet{dwork2010differential} considers a disease self-assessment website that monitors regional health conditions in real time while preserving the users' medical data privacy. \citet{farokhi2017smartgrid} design a smart meter that measures household energy consumption while preserving privacy against adversaries attempting to infer individual appliance usage patterns or the level of household occupancy. Similar practical examples can be seen in the fields of traffic flow estimation \citep{leny2014real}, image classification \citep{chen2024locally}, and online ads recommendation \citep{li2018DPDOL}, etc. 

Privacy mechanisms for system identification can be broadly categorized into encryption ones \citep{tan2023cooperative,yuan2025encryption} and stochastic obfuscation ones \citep{dwork2006calibrating,luo2025differential,li2025preserving}. 
Among them, the most renowned stochastic obfuscation mechanisms are differential privacy ones, introduced by \citet{dwork2006calibrating}. Subsequently, various stochastic obfuscation mechanisms based on information-theoretic metrics such as Fisher information \citep{farokhi2019ensuring}, conditional differential entropy \citep{wang2022decentralized}, and mutual information \citep{wang2016relation} have been developed. Compared to encryption mechanisms, stochastic obfuscation ones offer superior computational efficiency. Furthermore, their privacy level remains independent of the adversary's computational capabilities, rendering them resilient to potential threats such as quantum computing.

A fundamental issue in privacy-preserving system identification based on stochastic obfuscation is the trade-off between identification accuracy and privacy preservation \citep{cai2021cost}. In stochastic obfuscation mechanisms, excessive introduction of privacy noise to preserve measurement data can significantly degrade identification accuracy, whereas pursuing higher identification accuracy inevitably risks leakage of sensitive information. Therefore, under given privacy constraints, characterizing the fundamental limit of identification accuracy and designing corresponding algorithms that achieve optimal accuracy have become topics of widespread interest among researchers. Existing research primarily investigates this trade-off within pre-defined mechanisms, such as Gaussian \citep{asoodeh2018estimation,dong2022gaussian} or Laplacian noise ones \citep{dwork2009what,chen2024locally}, by adjusting noise levels. However, such trade-off analyses are only applicable to specific stochastic obfuscation mechanisms, and thereby lack generality. As a result, these analyses hardly provide guidance for the design of optimal stochastic obfuscation mechanisms and system identification algorithms.

To enhance the generality of the traditional trade-off results, this paper extends the classical Cram{\'e}r-Rao lower bound (CRLB) theory \citep{shao2003mathematical} to scenarios with privacy constraints, i.e., the privacy-preserving CRLB theory. 
Notably, beyond system identification algorithms, there are plenty of stochastic obfuscation mechanisms satisfying given privacy constraints, and the optimal design of system identification algorithms depends on the choice of stochastic obfuscation mechanisms. 
Consequently, the privacy-preserving CRLB theory is more challenging to establish compared with the classical one. 
This paper aims to overcome this challenge, characterizing a precise and attainable lower bound for identification accuracy applicable to general stochastic obfuscation mechanisms, and thereby establishing the fundamental trade-off between identification accuracy and privacy preservation.

%Therefore, this paper aims to characterize a precise and attainable lower bound for identification accuracy applicable to a broad class of stochastic obfuscation mechanisms. 

\subsection{Related Literature}

Trade-off between identification accuracy and privacy preservation has been extensively investigated in existing literature. For example, \citet{wangjm2023jssc} provide the trade-off analysis for the identification algorithm with polynomially damping privacy noise, and show that such an algorithm can achieve convergence under differential privacy constraint while the conventional approach with exponentially damping privacy noise \citep{huang2015DPOP} cannot. 
%proves that given differential privacy constraints, the algorithm with stochastic approximation step-sizes and polynomially damping privacy noise achieves superior identification accuracy compared to the conventional approach using exponentially damping step-sizes and privacy noise \cite{huang2015DPOP}. 
%compared with the algorithm with exponential damping step-size and privacy noise \cite{huang2015DPOP},  the identification accuracy is better under the algorithm with stochastic approximation step-size and polynomial damping privacy noise. 
%\cite{wangjm2023jssc} further points out that their algorithm with stochastic approximation step-size simultaneously achieves mean-square convergence and $\varepsilon$-differential privacy, with the mean squared error being inversely proportional to $\varepsilon^2$. 
\citet{wang2024tailoring,chen2024locally,ke2025privacy+quantization} extend the results to the constant and even increasing privacy noise case, and characterize the trade-off between convergence rate and privacy level of corresponding algorithms. Additionally, \citet{wang2024sample} improve the identification accuracy for given differential privacy constraint using time-varying sample sizes. More discussion for trade-off between identification accuracy and privacy preservation under specific algorithms can be seen in \citet{nozari2016functional,li2018DPDOL,saha2024privacy}. 

To broaden the applicability of trade-off analysis beyond specific algorithms, numerous works have focused on characterizing optimal additive privacy noise under different settings. \citet{geng2016optimal} derive the optimal noise distribution that minimizes the expected cost function under differential privacy constraint. \citet{dagan2022bounded} investigate the minimum bounded noise that can achieve a given differential privacy level. \citet{farokhi2019ensuring} employ Fisher information as a privacy metric and determine the noise distribution that achieves the optimal balance between identification accuracy and privacy preservation. 
The aforementioned works have extended traditional trade-off analysis from merely computing the optimal amplitude of privacy noise under a given distribution type to designing the optimal privacy noise distribution itself, which is important for the generalization of the trade-off analyses. 
However, these contributions still exhibit the following limitations. 
Firstly, these works are primarily confined to query-answering problems without measurement noises, precluding their application to system identification.
Additionally, by confining their scope to additive noise mechanisms, these works neglect the affine transformations often required for optimal trade-offs in system identification, alongside other privacy-preserving approaches such as multiplicative noise \citep{wang2022decentralized,brackenbury2019protecting} and quantizer-based mechanisms \citep{ke2025privacy+quantization,wang2023quantization}. 
%For more general system identification problems, it is often necessary to perform affine transformations on the data beyond simply adding privacy noise to achieve a better trade-off. 
%In addition to additive noise mechanisms and even the broader affine transformation mechanisms, there exist other privacy-preserving approaches such as multiplicative noise mechanisms \cite{wang2022decentralized,brackenbury2019protecting} and quantizer-based mechanisms \cite{ke2025privacy+quantization,wang2023quantization}. 

Some recent efforts have sought to address these limitations by developing privacy-preserving parameter identification algorithms that achieve optimal convergence rates under differential privacy constraints. 
\citet{cai2021cost,cummings2022mean} derive the mean squared error (MSE) lower bounds under centralized differential privacy constraints for general stochastic obfuscation mechanisms. The nearly optimal identification algorithms are also explicitly provided. 
On the other hand, \citet{duchi2019lower,barnes2020fisher} consider local differential privacy constraints, and investigate the lower bounds of identification accuracy by using communication complexity and Fisher information, respectively. 
Notably, these lower bounds on identification accuracy are not sufficiently tight. Most of these bounds involve unspecified constant factors, and in the absence of privacy constraints, these bounds do not degenerate to the classical CRLB. Therefore, the goal of this paper is to establish the precise lower bounds for identification accuracy under privacy constraints.
%Notably, all these lower bounds for identification accuracy involve unspecified constant factors, meaning that the actual estimation error of existing algorithms may deviate from the bound by a fixed multiplicative gap. Consequently, the aforementioned lower bounds are actually not precise, and the strict-sense attainability of such bounds remains unverified.  
%Therefore, the goal of this paper is to establish the precise and attainable lower bounds for identification accuracy under privacy constraints.  

\subsection{Main Contributions}

This paper establishes the privacy-preserving CRLB theory, characterizing a precise and attainable lower bound for identification accuracy applicable to general stochastic obfuscation mechanisms. 
%The main challenge lies in the fact that, without specifying stochastic obfuscation mechanisms or identification algorithms, 
%% when specific forms of privacy mechanisms and identification algorithms are not predetermined, 
%the domain of the lower bound derivation problem becomes extremely large, essentially leading to an infinite-dimensional optimization problem. Therefore, traditional analytical techniques based on parameter optimization \citep{wang2024sample} or even variational methods \citep{dagan2022bounded} become inapplicable. 
%This paper overcomes this difficulty within a unified framework where Fisher information simultaneously serves as the metric for privacy preservation \citep{farokhi2019ensuring,lip2025state} and the indicator of the identification accuracy bound \citep{barnes2020fisher,lip2025state}. The key lies in constructing an equivalent expression for the partial derivatives in the definition of Fisher information matrix, and then using the equivalent expression to establish an intrinsic relationship among identification accuracy, privacy level, and measurement noise.
%To overcome this challenge, this paper employs Fisher information to simultaneously characterize both the privacy level  \cite{farokhi2019ensuring,lip2025state} and the identification accuracy bound \cite{barnes2020fisher,lip2025state}, thereby creating a unified analytic framework for analyzing privacy-accuracy trade-offs. 
The main contributions are summarized as follows:

\begin{enumerate}[label={\roman*)}, leftmargin = 1em, itemsep=0pt, parsep=0pt,topsep=0pt,	partopsep=0pt]
	\item This paper derives an identifiability criterion under privacy constraint, and establishes the privacy-preserving CRLB in the identifiable case. The MSE of any admissible stochastic obfuscation mechanism with its corresponding unbiased identification algorithm is proven to be bounded below by this privacy-preserving CRLB. Different from the minimax lower bounds established in \citet{cai2021cost,cummings2022mean,duchi2019lower}, the proposed privacy-preserving CRLB is free of unspecified constant factors. Furthermore, it strictly reduces to the classical CRLB in the absence of privacy constraints, which validates its superior accuracy.
	
	\item Under Gaussian measurement noises, the privacy-preserving CRLB is strictly attainable. A Gaussian-mechanism-based privacy-preserving recursive least-squares (RLS) algorithm is formulated to guarantee that the estimates always attain privacy-preserving CRLB. In contrast, even under Gaussian measurement noises, the existing literature \citep{cai2021cost,cummings2022mean,duchi2019lower,barnes2020fisher} focuses solely on attainability in the sense of convergence rate.
	
	\item Under non-Gaussian measurement noises, the attainability of the privacy-preserving CRLB can also be established in the sense of  convergence rates. Specifically, a maximum-likelihood-based (ML-based) privacy-preserving identification algorithm is formulated, and it is proven that the gap between its MSE and the classical CRLB shares the same convergence rate as the difference between the privacy-preserving and classical CRLB. For comparison, \citet{cai2021cost} evaluate privacy cost by comparing their algorithm solely to the least-squares estimate, neglecting its MSE gap to the classical CRLB under non-Gaussian noise. \citet{cummings2022mean,duchi2019lower,barnes2020fisher} are limited to strong privacy regimes, and the estimation accuracy of their algorithms suffers from the usage of non-optimal algorithms, such as the direct averaging one.
	
	\item 
	The proposed privacy-preserving CRLB provides a unified benchmark for privacy-utility trade-off analysis across privacy mechanisms, system models, and privacy notions. Our results are applicable to general privacy mechanisms and identification algorithms, rather than specific ones \citep{wangjm2023jssc,wang2024tailoring,saha2024privacy}.
	Moreover, in contrast to the query-answering setups without measurement noise in \citet{geng2016optimal,dagan2022bounded,farokhi2019ensuring} and the mean estimation problems investigated in \citet{duchi2019lower,cai2021cost}, we consider the system identification problem with measurement noise and general measurement matrix. By establishing the relationship between Fisher information and differential privacy, the proposed lower bound can also be extended to the differential privacy constraint.  
\end{enumerate}

The rest of the paper is organized as follows.  Section \ref{sec:prob form} formulates the problem. Section \ref{sec:mainresults} presents the main results, including the identifiability analysis, the privacy-preserving CRLB, the corresponding recursive computation and attainability analysis. 
%Section \ref{sec:appl} shows the theoretical applicability. 
Section \ref{sec:simu} provides experimental examples to demonstrate the theoretical results. Section \ref{sec:concl} concludes the paper.

%\subsection*{Notation}

\textit{Notation.} In the rest of the paper, $\mathbb{N}$, $\mathbb{R}$, $\mathbb{R}^{n}$ and $\mathbb{R}^{m \times n}$ denote the sets of natural numbers, real numbers, $n$-dimensional real vectors and $m \times n$-dimensional real matrices, respectively.
%$\mathbb{I}_{\{\cdot\}}$ denotes the indicator function, whose value is 1 if its argument (a formula) is true, and 0, otherwise. 
$I_n$ is an $n\times n$ identity matrix. When the dimension is clear from the context, we simply write $I$ to denote an identity matrix of appropriate size. $ \diag\{\cdot\} $ denotes the block matrix formed in a diagonal manner of the corresponding matrices. $ \col\{\cdot\} $ denotes the column concatenation of the corresponding numbers or matrices.  $ \otimes $ denotes the Kronecker product. $A^+$ is the pseudo-inverse of $A$. Given symmetric matrices $A$ and $B$, $A \geq B$ (resp., $A>B$, $A\leq B$, and $A<B$) means that $A-B$ is positive semi-definite (resp., positive definite, negative semi-definite, and negative definite). $\mathcal{N}(\mu,\Sigma)$ is the Gaussian distribution with mean $\mu$ and covariance $\Sigma$. 

\section{Problem Formulation}\label{sec:prob form}

\subsection{System Model and Privacy Mechanisms}

The measurement system is given as follows. 
\begin{align}\label{sys}
	y=H\theta+w,
\end{align}
where $ y \in \mathbb{R}^{m} $ is the measurement, $ H \in \mathbb{R}^{m\times n} $ is the measurement matrix, $ w \in \mathbb{R}^{m} $ is the measurement noise, and $ \theta \in \mathbb{R}^{n} $ is the parameter to be identified. The measurement $y$ is the sensitive data to be preserved. 

\begin{rem}
	When $H$ is a column vector with all elements equal to $1$, and the components of the measurement noise  $w$ are independent and identically distributed and unbiased, the above privacy-preserving system identification problem degenerates to the classic private mean estimation problem \citep{cai2021cost,cummings2022mean}. 
%	Additionally, the system \eqref{sys} covers system identification issues under various communication models, including centralized identification with fusion centers \citep{barnes2020fisher}, distributed identification under networked communication \citep{wangjm2023jssc,chen2024locally}, as well as semi-distributed identification issues featuring both fusion centers and networked communication \citep{saha2024privacy}.
\end{rem}

For privacy, this paper primarily focuses on stochastic obfuscation mechanisms, exemplified by differential privacy mechanisms \citep{dwork2009what}. 
Here, we provide a generalized representation of such mechanisms. 
Each such mechanism can be characterized by a function $ M(\cdot,\cdot) $ and the density function $ f_d(\cdot) $ of the privacy noise. By employing the mechanism $ (M(\cdot,\cdot),f_d(\cdot)) $, a sensitive measurement $ y $ is transformed into a preserved output $ z = M(y,d) $. Here, $ d $ denotes the privacy noise, generated according to $ f_d(\cdot) $, which is independent of both the measurement noise $ w $ and the unknown parameter $ \theta $. Now, we define the admissible set $ \mathcal{C} $ for these stochastic obfuscation mechanisms as follows. 

\begin{defn}\label{def:admissible}
	\noindent{\normalfont\bfseries (Admissible Stochastic Obfuscation Mechanism).}\ For sensitive data \( y \in \mathbb{R}^m \), given a function \( M(\cdot,\cdot): \mathbb{R}^{m} \times \Omega_d \to \Omega_z \) and density function \( f_d(\cdot): \Omega_{d} \to \mathbb{R}_{\geq 0} \), the stochastic obfuscation mechanism \( (M(\cdot,\cdot),f_d(\cdot)) \) belongs to the admissible set \( \mathcal{C} \) if and only if  the conditional density or mass function \( p(z|y) \) is continuous and piecewise differentiable with respect to \(y\).
%	\begin{equation}\label{condi:admissible}
%		\mE\left[ \left(\frac{\partial  \ln p(z|d,\theta)}{\partial \theta} \right) \left( \frac{\partial \ln p(z|w,\theta)}{\partial \theta}\right)^\top \middle| \theta \right] = 0. 
%	\end{equation}
\end{defn}

%Due to the independence assumption between the privacy noise \( d \) and the measurement noise \( w \), t
The aforementioned admissible set \( \mathcal{C} \) includes a broad range of mechanisms. Please see Appendix \ref{appen:examples}.% \ref{subsec:app_mech}. 

\subsection{Stackelberg Game Perspective and Fisher Information Matrix Based Framework}\label{subsec:stackelberg}

This section models the privacy-preserving identification problem from the Stackelberg game perspective \citep{ungureanu2018stackelberg}. First, a potential adversary is introduced, who can design an identification algorithm $\hat{y}(z)$ to obtain an estimate of $y$. In this framework, the legitimate identifier acts as the leader of the Stackelberg game. Its objectives include minimizing its own identification error while ensuring that the adversary's identification accuracy regarding sensitive data does not exceed $S^{-1}$, where $S$ is the required privacy level. Specifically, the goal of the legitimate identifier can be formulated as:

\noindent \textbf{\quad Legitimate identifier:} 
\vspace{-0.5em} 
\begin{gather}\label{obj:id}
	\min_{\hat{\theta}(\cdot),M(\cdot,\cdot)} \mathbb{E} \left[ \left(\hat{\theta}(z) - \theta\right)\left(\hat{\theta}(z) - \theta\right)^\top \middle| \theta \right] 
\end{gather}
\vspace{-2.2em}
\begin{align*}
	\quad \text{s.t.\ }& z = M(y,d),\ 
	\mathbb{E} \left[ \left(\hat{y}(z) - y\right)\left(\hat{y}(z) - y\right)^\top \middle| y \right] \geq S^{-1}.
\end{align*}
The potential adversary, acting as a follower who has the full knowledge of the adopted privacy mechanism \( (M(\cdot,\cdot),f_d(\cdot)) \), aims to minimize the estimation error of $y$, which can be formulated as:

\noindent \textbf{\quad Potential adversary:}
\vspace{-0.5em} 
\begin{gather}\label{obj:adv}
	\min_{\hat{y}(\cdot)} \mathbb{E} \left[ \left(\hat{y}(z) - y\right)\left(\hat{y}(z) - y\right)^\top \middle| y \right] 
\end{gather}
\vspace{-1em}

In the aforementioned problem modeling, it can be observed that given a fixed privacy mechanism, the optimization objectives for both the legitimate identifier and the potential attacker degenerate into optimal identification problems. Consequently, the classical CRLB theory can be introduced to investigate this issue. 

\begin{prop}\label{prop:Ocrbound}
	\noindent{\normalfont\bfseries (\citet{shao2003mathematical}, Theorem 3.3).}\ Given output $z$ and parameter or sensitive data $ y $, if $ \mathcal{I}_z(y) $ is invertible, then for any unbiased estimator $ \hat{y} = \hat{y}(z) $ of $ y $, $ \mE(\hat{y}-y)(\hat{y}-y)^\top \geq \mathcal{I}_z^{-1}(y) $, where
	\begin{align}\label{def:Fisher}
		\mathcal{I}_z(y)
		= \mE \left[ \left[\frac{\partial \ln(p(z|y))}{\partial y}\right]\left[\frac{\partial \ln(p(z|y))}{\partial y}\right]^\top \middle| y \right]. 
	\end{align}
\end{prop}

\begin{defn}
	\noindent{\normalfont\bfseries (Fisher information matrix, \citet{shao2003mathematical,zamir1998Fisher}).}\ Fisher information matrix of output $ z $ with respect to parameter or sensitive data $ y $ is defined as \eqref{def:Fisher}. 
	Given a random variable $ x $, the conditional Fisher information matrix is defined as  
	\begin{align*}
		\mathcal{I}_z(y|x)
		= \mE \left[ \left[\frac{\partial \ln(p(z|x,y))}{\partial y}\right]\left[\frac{\partial \ln(p(z|x,y))}{\partial y}\right]^\top \middle| y \right].
	\end{align*}
\end{defn}

Within the framework of \eqref{obj:id} and \eqref{obj:adv}, CRLB and Fisher information quantify the fundamental limits of identification for both legitimate identifier and potential adversary. This allows the original problem to be reformulated as the following privacy-preserving CRLB determination task, where Fisher information matrix serves as the privacy metric \citep{farokhi2019ensuring,lip2025state}. 

\begin{defn}
	\noindent{\normalfont\bfseries (Privacy-Preserving CRLB).}\ Given the privacy constraint \( \mathcal{I}_z (y) \leq S \), a positive semidefinite matrix \( R_S \in \mathbb{R}^{n\times n} \) is a privacy-preserving CRLB if 
	\begin{align}\label{def:RS}
		\mE\left[ \left(\hat{\theta}(z) - \theta\right)\left(\hat{\theta}(z) - \theta\right)^\top \middle| \theta \right] \geq R_S
	\end{align}
	for any admissible stochastic obfuscation mechanism \( (M(\cdot,\cdot),f_d(\cdot)) \) satisfying this privacy constraint and any unbiased estimate \( \hat{\theta} (z) \) of the unknown parameter \( \theta \).
\end{defn}

%Different from our Fisher-information-matrix-based framework, \citet{cai2021cost,cummings2022mean} adopt differential privacy as privacy metrics for privacy-preserving mean estimation. 
Fisher information matrix is highly related to differential privacy \citep{cai2021cost,cummings2022mean}. 
%Here, we introduce the definitions of event-level and user-level differential privacy \cite{dwork2010differential} and provide the theoretical characterization of the relation between Fisher information and differential privacy. Since the definitions for both single-user and multi-user scenarios are equivalent, this section focuses only on the single-user case for simplicity.

\begin{defn}\label{def:DP}
	\noindent{\normalfont\bfseries (Differential privacy, \citet{dwork2006calibrating}).}\ Given $\varepsilon> 0$, a stochastic obfuscation mechanism $(M(\cdot,\cdot),f_d(\cdot))$ with output sequence $ \{z_k\} $ is $\varepsilon$-differentially private if for any measurable set $\mathcal{O}$ and all sensitive sequences $\{y_k\}$ and $\{y_k^\prime\}$ that differ in time $t$,
	\begin{align}\label{ineq:def_DP}
		\mP( \{z_k\} \in \mathcal{O}|\{y_k\})
		\leq \mP( \{z_k\} \in \mathcal{O}|\{y_k^\prime\}) e^{\varepsilon \lVert y_t - y_t^\prime \rVert}.
	\end{align}
%	In the single-user case, the mechanism
%	$(M(\cdot,\cdot),f_d(\cdot))$ is \textit{user-level} $\varepsilon$-differential privacy if for any measurable set $\mathcal{O}$ and all $\{y_k\}$ and $\{y_k^\prime\}$ that differ in \textit{all} $k$, \eqref{ineq:def_DP} holds. 
\end{defn}

\begin{rem}
	Compared with the classical differential privacy \citep{dwork2010differential}, we introduce a term $  \lVert y_t - y_t^\prime \rVert$ in \eqref{ineq:def_DP} to reflect the fact that indistinguishability increases as the two adjacent sensitive sequences get closer. Since the privacy level is linearly correlated with  $ \lVert y_t - y_t^\prime \rVert$ in most classical differential-privacy-based studies \citep{chen2024locally,cai2021cost,wang2024tailoring}, Definition \ref{def:DP} remains consistent with existing literature.
\end{rem}

\begin{prop}\label{prop:DP2Fisher}
	For  $ Y = \col\{y_1, \ldots, y_K\} $, $\varepsilon$-differential privacy implies 
	$ \trace(\mathcal{I}_{z}(Y)) \leq K \varepsilon^2. $
%	\begin{enumerate}[label={\alph*)},leftmargin=1.4em]
%		\item Event-level $\varepsilon$-differential privacy implies 
%		$$ \trace(\mathcal{I}_{z}(Y)) \leq K \varepsilon^2; $$
%		\item User-level $\varepsilon$-differential privacy implies 
%		$$ \trace(\mathcal{I}_{z}(Y)) \leq \varepsilon^2, $$ 
%	\end{enumerate}
%	where.
\end{prop}

\begin{pf}
%	Under event-level $\varepsilon$-differential privacy
	Without loss of generality,
	assume that adjacent $Y$ and $Y^\prime$ differ for only $y_k$ and $y_k^\prime$. Denote $ \delta = \Absl{y_k - y_k^\prime} $ and $\Delta y_k = \frac{1}{\delta} (y_k - y_k^\prime)$. Then, $y_k - y_k^\prime = \delta \Delta y_k $. Therefore, by Lagrange mean value theorem \citep{zorich2015math} and the definition of differential privacy, 
	\begin{align*}
		\ln p\left( z \middle| Y \right) - \ln p\left( z \middle| Y^\prime \right)
		= \delta \frac{\partial \ln p\left( z \middle| Y \right)}{\partial y_k} \Delta y_k + O(\delta^2) 
		\leq \varepsilon \delta. 
	\end{align*}
	Letting $\delta \to 0$, we have $\frac{\partial \ln p\left( z \middle| Y \right)}{\partial y_k} \Delta y_k \leq \varepsilon$.
	%	\begin{align*}
		%		\frac{\partial \ln p\left( z \middle| Y \right)}{\partial y_k} \Delta y_k \leq \varepsilon. 
		%	\end{align*}
	Note that $\Delta y_k$ can be any unit vector. Then, we have 
	$\Abs{\frac{\partial \ln p\left( z \middle| Y \right)}{\partial y_k}} \leq \varepsilon$, 
	which implies that
	\begin{equation*}
		\trace(I_z(Y)) = \sum_{k=1}^{K} \mE\left[ \Abs{\frac{\partial \ln p\left( z \middle| Y \right)}{\partial y_k}}^2 \middle| Y \right]
		\leq K \varepsilon^2. \hfill \qed
	\end{equation*}
%	\hfill $\qed$
\end{pf}

%\begin{rem}
%	\cite{barnes2020fisher} also studies the relation between differential privacy and Fisher information. However, it adopts a stronger adjacency notion where all sensitive data values are mutually adjacent, which deviates from standard event-level or user-level definitions and thus makes it difficult to accurately characterize the actual impact of classical differential privacy on estimation. 
%	In contrast, Proposition \ref{prop:DP2Fisher} establishes this relation under Definition \ref{def:DP}. Moreover from Proposition \ref{prop:DP2Fisher}, one can gain deeper insights into the differences between event-level and user-level differential privacy from the perspective of Fisher information.
%\end{rem}

\subsection{Problem Statement}

Now, we address the following three goals for this paper:

\begin{enumerate}[leftmargin = 1.5em,label={\rm \roman*)}]
	\item \textbf{Identifiability analysis:} Establish an identifiability criterion for measurement system \eqref{sys} under the privacy constraint \( \mathcal{I}_z (y) \leq S \).
	\item \textbf{Lower bound computation:} Build the privacy-preserving CRLB under the identifiable condition, and provide recursive computation formulas for the multi-measurement systems.
	\item \textbf{Attainability analysis:} Develop  identification algorithms that can attain the privacy-preserving CRLB.
\end{enumerate}

\section{Main Results}\label{sec:mainresults}

\subsection{Identifiability Analysis Under Privacy Constraint}\label{subsec:identifiable}

This subsection analyzes the identifiability of measurement system \eqref{sys} under the privacy constraint \( \mathcal{I}_z (y) \leq S \). The analysis relies on the following assumption regarding the measurement noise \( w \).

\begin{assum}\label{assum:w0}
	For the measurement noise \( w \), the density function $f_w(\cdot)$ is continuous and piecewise differentiable, and \( \mathcal{I}_w(\theta) = 0 \). 
\end{assum}

\begin{rem}
	Assumption \ref{assum:w0} indicates that $w$ is purely random noise, and guarantees that observing the noise will not help the identification process.
\end{rem}

Based on the Fisher information matrix, we now define identifiability under privacy constraint as below. 

\begin{defn}\label{def:identifiable}
	\noindent{\normalfont\bfseries (Identifiability Under Privacy Constraint).}\ Given positive semi-definite \( S \), if there exists an admissible stochastic obfuscation mechanism \( (M(\cdot,\cdot),f_d(\cdot)) \) satisfying the privacy constraint \( \mathcal{I}_z (y) \leq S \) such that \( \mathcal{I}_{z}(\theta) > 0 \), then the measurement system \eqref{sys} is called identifiable under privacy level \( S \).
\end{defn}

\begin{thm}\label{thm:identifiable}
	Under Assumption \ref{assum:w0}, the measurement system \eqref{sys} is identifiable under privacy level \( S \) if and only if the matrix $ H^\top S H $ is invertible. 
\end{thm}

\begin{pf}
	\textbf{Necessity.} By Corollary A.1 of \citet{ke2025privacy+quantization} and Assumption \ref{assum:w0}, 
	\begin{align*}
		\mathcal{I}_z (\theta) & \leq \mathcal{I}_z(\theta|w) \\
		= & \mE \left[ \left[\frac{\partial \ln p(z|w,\theta)}{\partial \theta}\right]\left[\frac{\partial \ln p(z|w,\theta)}{\partial \theta}\right]^\top \middle| \theta \right] \\
		= & H^\top \mE \left[ \left[\frac{\partial \ln p(z|y)}{\partial y}\right]\left[\frac{\partial \ln p(z|y)}{\partial y}\right]^\top \middle| \theta \right] H \\
		= &  H^\top \mathbb{E} [\mathcal{I}_z (y)| \theta] H \leq H^\top S H. 
	\end{align*}
	Therefore, \( \mathcal{I}_z (\theta) > 0 \) implies \( H^\top S H > 0 \). 
	
	\textbf{Sufficiency.} When \( H^\top S H \) is invertible, define the stochastic obfuscation mechanism
	\begin{align}\label{eq:mech}
		z = M(y,d) = S y + d,
	\end{align}
	where \( d \sim \mathcal{N}(0,S) \). Then, $ \mathcal{I}_z (y) = S $. 
	Besides, substituting \eqref{sys} into \eqref{eq:mech}, we have $z = S H \theta + S w + d $. 
	
	Consider $z^\prime = z + d^\prime $ with $d^\prime \sim \mathcal{N}(0,I_m)$, which is independent of $w$ and $d$. Then by Lemma 1 of \citet{zamir1998Fisher}, $\mathcal{I}_{z} (\theta) \geq \mathcal{I}_{z^\prime} (\theta) $. Therefore, it suffices to prove $\mathcal{I}_{z^\prime} (\theta) > 0$. 
	
	Define $w_S = S w + d + d^\prime$. Then, since $\mP\{w_S^\top u = c \} < 1 $ for all constant $c$, we have
	$u^\top \nabla f_{w_S} (w_S) \neq 0 $ for some $w_S$. Hence by the definition of Fisher information, $ u^\top \mathcal{I}_z(SH\theta) u > 0$. Therefore, we have $u^\top \mathcal{I}_{z^\prime} (S H \theta) u > 0$ for all unit vector $u$, which implies that $ \mathcal{I}_{z^\prime}(S H \theta) > 0$. 
	
	By Lemma 4 of \citet{zamir1998Fisher}, we have
	\begin{align*}
		\mathcal{I}_{z^\prime}(\theta) =  H^\top S \mathcal{I}_{z^\prime}(S H \theta) S H > \lambda_{\min} (\mathcal{I}_{z^\prime}(S H \theta)) H^\top S^2 H.
	\end{align*}
%	Therefore, it suffices to prove that $H^\top S^2 H$ is invertible. 
	By the invertibility of $H^\top S H $, and 
	\begin{align*}
		\rank(H^\top S H) \leq \rank(S H) = \rank (H^\top S^2 H), 
	\end{align*}
	one can get $H^\top S^2 H$ is also invertible. Therefore, $\mathcal{I}_{z^\prime}(\theta) > 0$, which implies $\mathcal{I}_{z}(\theta) > 0$. 
	
%	then $u^\top \mathcal{I}_z(S H \theta) u \to \infty$. Otherwise, 
%	 and the unbiased estimate 
%	\begin{align*}
%		\hat{\theta}(z) = \left(H^\top S H\right)^{-1} H^\top S^{\frac{1}{2}} z. 
%	\end{align*}
%	Since \( d \) follows a Gaussian distribution, by Lemma 4 of \cite{zamir1998Fisher}, we have \( \mathcal{I}_z(y) = S^{\frac{1}{2}} I_m S^{\frac{1}{2}} = S \), satisfying the privacy constraint.
%	
%	Furthermore, since
%	\begin{align*}
%		\hat{\theta}(z) = &  \left(H^\top S H\right)^{-1} H^\top S^{\frac{1}{2}} z \\
%		= & \left(H^\top S H\right)^{-1} H^\top S^{\frac{1}{2}} \left( S^{\frac{1}{2}} \left(y - \mE w \right) + d \right) \\
%		= &  \left(H^\top S H\right)^{-1} H^\top S \left(y - \mE w \right) + \left(H^\top S H\right)^{-1} H^\top S^{\frac{1}{2}} d \\
%		= & \theta + \left(H^\top S H\right)^{-1} H^\top S \left( w - \mE w \right) \\
%		& + \left(H^\top S H\right)^{-1} H^\top S^{\frac{1}{2}} d,
%	\end{align*}
%	it follows that
%	\begin{align*}
%		& \mE \left[ \left( \hat{\theta}(z) - \theta \right)  \left( \hat{\theta}(z) - \theta \right)^\top \middle| \theta \right] \\
%		= & \left(H^\top S H\right)^{-1} H^\top S \Var(w) S H \left(H^\top S H\right)^{-1} + \left(H^\top S H\right)^{-1} \\
%		< & \infty. 
%	\end{align*}
%	Thus, by Theorem \ref{thm:Ocrbound}, \( \mathcal{I}_{z}(\theta) > 0 \).
	
	In conclusion, the measurement system \eqref{sys} is identifiable under privacy level \( S \).
\hfill $\qed$
\end{pf}

\begin{rem}
	When the privacy level $S$ is positive definite, the identification criterion in Theorem \ref{thm:identifiable} is equivalent to the condition that $H^{\top} H$ is invertible, which is exactly the identification criterion without considering privacy. 
%	Furthermore, according to Theorem \ref{thm:identifiable}, for certain measurement matrices \( H \), the identifiability criterion can hold even when $S$ is not invertible. This is because when a row vector in \( H \) can be spanned by other row vectors, even if the corresponding component of measurement \( y \) is subject to perfect privacy constraint and therefore not transmitted at all, the information from that component can still be compensated through other components. 
\end{rem}

\subsection{Privacy-Preserving CRLB}\label{subsec:crbound}

This subsection aims to analyze the privacy-preserving CRLB and the privacy-preserving Fisher information matrix. 

\begin{thm}\label{thm:crbound}
	\noindent{\normalfont\bfseries (Privacy-Preserving CRLB).}\ Under Assumption \ref{assum:w0}, given the measurement system \eqref{sys} and privacy constraint \( \mathcal{I}_z (y) \leq S \), if \( H^\top S H \) is invertible, then for any admissible stochastic obfuscation mechanism \( (M(\cdot,\cdot),f_d(\cdot)) \) satisfying this privacy constraint and any unbiased estimate \( \hat{\theta}(z) \), we have
	\begin{align}\label{concl:PPCRB}
		\mE\left[ \left(\hat{\theta}(z) - \theta\right)\left(\hat{\theta}(z) - \theta\right)^\top \middle| \theta \right] \geq \Sigma_{\text{\rm PPCR}}, 
	\end{align} 
	where
	\begin{align*}
		\Sigma_{\text{\rm PPCR}} = \left(H^\top S^{\frac{1}{2}} \left( S^{\frac{1}{2}} \left(\mathcal{I}_{y}(H\theta)\right)^{-1} S^{\frac{1}{2}} + I_m \right)^{-1} S^{\frac{1}{2}} H\right)^{-1}.
	\end{align*}
%	Furthermore, if the measurement noise \( w \) follows a Gaussian distribution \( \mathcal{N}(\mu_w,\Sigma_w) \), then when 
%	\begin{align}
%		&z = M(y,d) = S^{\frac{1}{2}} \left(y - \mu_w \right) + d,\ d \sim \mathcal{N}(0,I_m), \label{optM}\\
%		&\hat{\theta}(z) = \Sigma_{\text{\rm PPCR}} H^\top  S^{\frac{1}{2}} \! \left(  S^{\frac{1}{2}} \Sigma_w S^{\frac{1}{2}} \! + \! I_m  \right)\!^{-1} \! \left( z \! -  S^{\frac{1}{2}} \mu_{w} \right)\!, \label{optEst}
%	\end{align}
%	\eqref{concl:PPCRB} holds with equality. In this case, the privacy-preserving Fisher information matrix \( \mathcal{PI}_{y|S} (\theta) = \Sigma_{\text{\rm PPCR}}^{-1} \).
\end{thm}

\begin{pf}
	By Proposition \ref{prop:Ocrbound}, \eqref{concl:PPCRB} is equivalent to proving that for any admissible stochastic obfuscation mechanism \( (M(\cdot,\cdot),f_d(\cdot)) \) satisfying \( \mathcal{I}_z (y) \leq S \), we have
	\begin{align*}
		\mathcal{I}_z (\theta) \leq H^\top S^{\frac{1}{2}} \left( S^{\frac{1}{2}}  \left(\mathcal{I}_{y}(H\theta)\right)^{-1} S^{\frac{1}{2}} + I_m \right)^{-1}   S^{\frac{1}{2}} H. 
	\end{align*}
	Furthermore, by Lemma 4 of \citet{zamir1998Fisher}, it suffices to prove
	\begin{align}\label{ineq:pre/ppcr}
		\mathcal{I}_z (H\theta) \leq S^{\frac{1}{2}} \left( S^{\frac{1}{2}}  \left(\mathcal{I}_{y}(H\theta)\right)^{-1} S^{\frac{1}{2}} + I_m \right)^{-1}   S^{\frac{1}{2}}. 
	\end{align}
	
	Note that  \( p(z|H\theta) = \int_{w\in \mathbb{R}^m} p(z|w,H\theta) p(w)  \text{d} w \), which implies $\frac{\partial p(z|H\theta)}{\partial H\theta} 
	= \int_{w\in \mathbb{R}^m} \frac{\partial p(z|w,H\theta)}{\partial H\theta} f_w(w)  \text{d} w$.
%	\begin{align*}
%		\frac{\partial p(z|H\theta)}{\partial H\theta} 
%		= \int_{w\in \mathbb{R}^m} \frac{\partial p(z|w,H\theta)}{\partial H\theta} f_w(w)  \text{d} w. 
%	\end{align*}
	Then,
	\begin{align}\label{eq:lnp=Ew}
		&\frac{\partial  \ln p(z|H \theta)}{\partial H\theta}
		= \frac{\frac{\partial p(z|H\theta)}{\partial H\theta}}{p(z|H\theta)} 
		= \int_{w\in \mathbb{R}^m} \frac{\frac{\partial p(z|w,H\theta)}{\partial H\theta}}{p(z|H\theta)} f_w(w)  \text{d} w \nonumber\\
		=& \int_{w\in \mathbb{R}^m} \frac{\frac{\partial p(z|w,H\theta)}{\partial H\theta}}{p(z|w,H\theta)} \frac{f_w(w)p(z|w,H\theta)}{p(z|H\theta)}  \text{d} w \nonumber\\
		=& \int_{w\in \mathbb{R}^m} \frac{\frac{\partial p(z|w,H\theta)}{\partial H\theta}}{p(z|w,H\theta)} p(w|z,H\theta)  \text{d} w\nonumber\\
		=& \mE_w \left[ \frac{\frac{\partial p(z|w,H\theta)}{\partial H\theta}}{p(z|w,H\theta)} \middle| z,H\theta\right] 
		=  \mE_w \left[ \frac{\partial  \ln p(z|w,H\theta)}{\partial H\theta} \middle| z,H\theta\right]. 
	\end{align}
	Additionally, since $	p(z|H\theta) 
	= \int_{y\in \mathbb{R}^m} p(z|y) p(y|H\theta)  \text{d} y
	= \int_{y\in \mathbb{R}^m} p(z|y) f_w(y-H\theta)  \text{d} y$,
%	\begin{align*}
%		p(z|H\theta) 
%		= \int_{y\in \mathbb{R}^m} p(z|y) p(y|H\theta)  \text{d} y
%		= \int_{y\in \mathbb{R}^m} p(z|y) f_w(y-H\theta)  \text{d} y,
%	\end{align*}
	which implies
	\begin{align*}
		\frac{\partial p(z|H\theta)}{\partial H\theta} 
		=& \int_{y\in \mathbb{R}^m} p(z|y) 
		\frac{\partial f_w(y-H\theta)}{\partial H\theta}   \text{d} y\\ 
		=& - \int_{y\in \mathbb{R}^m} p(z|y) 
		\frac{\partial f_w(w)}{\partial w}   \text{d} y
	\end{align*}
	Then, 
	\begin{align}\label{eq:lnp=Ey}
		&\frac{\partial  \ln p(z|H \theta)}{\partial H\theta}
		= \frac{\frac{\partial p(z|H\theta)}{\partial H\theta}}{p(z|H\theta)} 
		= - \int_{y\in \mathbb{R}^m} p(z|y) \frac{\frac{\partial f_w(w)}{\partial w} }{p(z|H\theta)} \text{d} y \nonumber\\
		= & - \int_{y\in \mathbb{R}^m} \frac{p(z|w,H\theta) f_w(w)}{p(z|H\theta)} \frac{\frac{\partial f_w(w)}{\partial w} }{f_w(w)} \text{d} y \nonumber\\
%		= & - \int_{y\in \mathbb{R}^m} p(w|z,H\theta)\frac{\frac{\partial f_w(w)}{\partial w}}{f_w(w)} \text{d} y \nonumber\\
		= & - \int_{w\in \mathbb{R}^m} p(w|z,H\theta) \frac{\frac{\partial f_w(w)}{\partial w}}{f_w(w)} \text{d} w \nonumber\\
		= & - \mE_w \left[\frac{\frac{\partial f_w(w)}{\partial w}}{f_w(w)} \middle| z,H\theta\right] 
		= - \mE_w \left[ \frac{\partial \ln f_w(w)}{\partial w} \middle| z,H\theta\right]. 
	\end{align}
	
	Given any matrices \( A, B \in \mathbb{R}^{m\times m} \), by \eqref{eq:lnp=Ew} and \eqref{eq:lnp=Ey}, 
	\begin{align*}
		& (A+B) \left[ \frac{\partial \ln p(z|H\theta)}{\partial H\theta} \right] \left[ \frac{\partial \ln p(z|H\theta)}{\partial H\theta} \right]^\top (A+B)^\top \\
		= &  \mE\left[ A \frac{\partial  \ln p(z|w, H\theta)}{\partial H\theta} - B \frac{\partial \ln f_w(w)}{\partial w} \middle| z,H\theta \right] \nonumber\\
		& \cdot \mE\left[ A \frac{\partial  \ln p(z|w, H\theta)}{\partial H\theta} - B \frac{\partial \ln f_w(w)}{\partial w} \middle|z, H\theta \right]^\top,
	\end{align*}
	where
	\begin{align*}
		& \mE\left[ \frac{\partial  \ln p(z|w, H\theta)}{\partial H\theta} \middle| z,H\theta \right] \mE\left[ \frac{\partial  \ln p(z|w, H\theta)}{\partial H\theta} \middle| z,H\theta \right]^\top, \\
		\leq & \mE\left[ \left(\frac{\partial  \ln p(z|w, H\theta)}{\partial H\theta}\right)  \left(\frac{\partial  \ln p(z|w, H\theta)}{\partial H\theta}\right)^\top \middle| z,H\theta \right]
	\end{align*}
	\begin{align*}
		& \mE\left[ \frac{\partial \ln f_w(w)}{\partial w} \middle| z,H\theta \right] \mE\left[ \frac{\partial \ln f_w(w)}{\partial w} \middle| z,H\theta \right]^\top \\
		\leq & \mE\left[ \left(\frac{\partial \ln f_w(w)}{\partial w}\right)  \left(\frac{\partial \ln f_w(w)}{\partial w}\right)^\top \middle| z,H\theta \right]
	\end{align*}
	Furthermore, by Lemma \ref{lemma:PPCR_S} in Appendix \ref{appen:proof}, 
	\begin{align*}
		& \mE \left[\left(\frac{\partial  \ln p(z|w,H\theta)}{\partial H\theta}\right) \left( \frac{\partial \ln f_w(w)}{\partial w} \right)^\top \middle| H\theta\right] \nonumber\\
		= & \mE \left[ \mE \! \left[ \frac{\partial  \ln p(z|w,H\theta)}{\partial H\theta} \middle| w, H\theta\right] \! \left( \frac{\partial \ln f_w(w)}{\partial w} \right)^\top \middle| H\theta\right] \! = 0. 
	\end{align*}
	Therefore, 
	\begin{align*}
		& (A+B) \mathcal{I}_{z}(H\theta)  (A+B)^\top \nonumber\\
		= & \mE \left[ (A+B) \left[ \frac{\partial \ln p(z|H\theta)}{\partial H\theta} \right]\right. \\
		& \qquad \left. \cdot \left[ \frac{\partial \ln p(z|H\theta)}{\partial H\theta} \right]^\top  (A+B)^\top \middle| H\theta\right] \nonumber\\
		\leq & A \mE\left[ \left(\frac{\partial  \ln p(z|w, H\theta)}{\partial H\theta}\right)  \left(\frac{\partial  \ln p(z|w, H\theta)}{\partial H\theta}\right)^\top \middle| H\theta \right] A^\top \nonumber\\
		& + B \mE\left[ \left(\frac{\partial \ln f_w(w)}{\partial w}\right)  \left(\frac{\partial \ln f_w(w)}{\partial w}\right)^\top \middle| H\theta \right] B^\top \nonumber\\
		= & A \mathcal{I}_z(H\theta|w) A^\top + B \mathcal{I}_y (H\theta) B^\top, 
	\end{align*}
	which together with 
	\begin{align}\label{ineq:ES}
		& \mathcal{I}_z (H\theta|w) \nonumber\\
		= & \mE \left[ \left[\frac{\partial \ln p(z|w, H\theta)}{\partial H\theta}\right]\left[\frac{\partial \ln p(z|w, H\theta)}{\partial H\theta}\right]^\top \middle| H\theta \right] \nonumber\\
		= &\mE \left[ \left[\frac{\partial \ln p(z|y)}{\partial y}\right]\left[\frac{\partial \ln p(z|y)}{\partial y}\right]^\top \middle| H\theta \right] 
		= \mathbb{E} [\mathcal{I}_z (y)| H\theta] \nonumber\\
		\leq & S 
	\end{align}
	implies $(A+B) \mathcal{I}_{z}(H\theta)  (A+B)^\top
	\leq A S A^\top + B \mathcal{I}_{y}(H\theta) B^\top$.
%	\begin{align*}
%		(A+B) \mathcal{I}_{z}(H\theta)  (A+B)^\top
%		\leq A S A^\top + B \mathcal{I}_{y}(H\theta) B^\top.
%	\end{align*}
	
	For any \( \epsilon > 0 \), define \( S_\epsilon = (S^{\frac{1}{2}} + \epsilon I_m)^2 \), which satisfies \( S_\epsilon > S \) and is invertible. Then,
	\begin{align*}
		(A+B) \mathcal{I}_{z}(H\theta)  (A+B)^\top
		\leq A S_\epsilon A^\top + B \mathcal{I}_{y}(H\theta) B^\top.
	\end{align*}
	Set \( A = S_\epsilon^{-1}, B = \left(\mathcal{I}_{y}(H\theta)\right)^{-1} \). Then,
	\begin{align*}
		& \mathcal{I}_{z}(H\theta)
		\leq \left(S_\epsilon^{-1} + \left(\mathcal{I}_{y}(H\theta)\right)^{-1} \right)^{-1} \\
		= & \left( S^{\frac{1}{2}} + \epsilon I_m \right) \left(\left( S^{\frac{1}{2}} + \epsilon I_m \right) \right. \\ 
		& \cdot \left. \left(\mathcal{I}_{y}(H\theta)\right)^{-1} \left( S^{\frac{1}{2}} + \epsilon I_m \right) + I_m \right)^{-1}  \left( S^{\frac{1}{2}} + \epsilon I_m \right).
	\end{align*}
	Since this inequality holds for all \( \epsilon > 0 \) and the right-hand side is continuous, \eqref{ineq:pre/ppcr} holds, implying \eqref{concl:PPCRB}.
\hfill $\qed$
\end{pf}

%\begin{rem}
%	Theorem \ref{thm:crbound} establishes the privacy-preserving CRLB. 
%	The key is to use \eqref{eq:lnp=Ew} and \eqref{eq:lnp=Ey} to link $\mathcal{I}_z(H\theta)$, $\mathcal{I}_z(y)$, and $\mathcal{I}_y(H\theta)$ through a mechanism-independent matrix inequality.
%%	The key is using \eqref{eq:lnp=Ew} and \eqref{eq:lnp=Ey} to establish the equivalent relation between $\mathcal{I}_z(H\theta)$, $\mathcal{I}_z(y)$ and $\mathcal{I}_y(H\theta)$. 
%\end{rem}

\begin{rem}
	According to \eqref{ineq:ES}, the privacy-preserving CRLB also holds under the privacy constraint $ \mE \left[ \mathcal{I}_z(y) \middle| \theta \right] \leq S$. However, satisfying only this averaged constraint implies that $\mathcal{I}_z(y)$ might take significantly large values for certain realizations of $y$, thereby exposing $y$ to the risk of being estimated with high precision. Therefore, this paper primarily considers the stricter pointwise privacy constraint $\mathcal{I}_z(y)\leq S$.  
\end{rem}

\begin{rem}
	Theorem \ref{thm:crbound} can be extended to the nonlinear measurement system $ y = F(\theta) + w $ by replacing $ H $ with $ F^\prime (\theta) $ and $ \mathcal{I}_{y}(H\theta) $ with $\mathcal{I}_{y}(F(\theta))$. The nonlinear result follows directly by substituting \(F(\theta)\) for \(H\theta\) in the proof.
%	This is because $H\theta$ is treated as a single entity throughout the proof of \eqref{ineq:pre/ppcr}. Consequently, substituting $F(\theta)$ for $H\theta$ in the nonlinear case does not affect the validity of the proof. By further applying Lemma 4 of \citet{zamir1998Fisher}, the privacy-preserving CR bound for the nonlinear system can be derived.
\end{rem}

The following corollary characterizes the gap between the privacy-preserving CRLB and the classical one $\Sigma_{\text{\rm CR}} = \left(\mathcal{I}_{y}(\theta)\right)^{-1} $.

\begin{cor}\label{coro:PPCR}
	Under the condition of Theorem \ref{thm:crbound}, we have
	$\Sigma_{\text{\rm PPCR}} \geq \Sigma_{\text{\rm CR}} + \left(H^\top S H\right)^{-1}$.
%	\begin{align*}
%		\Sigma_{\text{\rm PPCR}} \geq \Sigma_{\text{\rm CR}} + \left(H^\top S H\right)^{-1}. 
%	\end{align*}
\end{cor}

The proof is provided in Appendix \ref{appen:proof}.

By applying Corollary \ref{coro:PPCR}, we can get the estimation lower bounds under differential privacy. 

\begin{cor}\label{coro:DP}
	Consider the multi-measurement system 
	\begin{align}\label{sys:mult}
		y_k = H_k \theta + w_k,\ k = 1,\ldots,K, 
	\end{align}
	with independent $\{w_k\}$ following Assumption \ref{assum:w0} and invertible $\sum_{k=1}^K H_k^\top H_k$.  Then, for unbiased estimator $\hat{\theta}(z)$, 
	\begin{equation*}
		\mE \! \left[ \! \left(\hat{\theta}(z) - \theta\right) \! \left(\hat{\theta}(z) - \theta\right)^\top \middle| \theta \right] \! \geq \Sigma_{\text{\rm CR}} + \frac{\left(\sum_{k=1}^K H_k^\top H_k\right)^{-1}}{\trace(\mathcal{I}_z(\bar{Y}_K))}, 
	\end{equation*}
	where $\bar{Y}_K = \col\{y_1,\ldots,y_K\}$. 
	Therefore, under $\varepsilon$-differential privacy constraint, 
	\begin{equation*}
		\mE \! \left[ \! \left(\hat{\theta}(z) - \theta\right) \! \left(\hat{\theta}(z) - \theta\right)^\top \middle| \theta \right] \! \geq \Sigma_{\text{\rm CR}} + \frac{\left(\sum_{k=1}^K H_k^\top H_k\right)^{-1}}{K\varepsilon^2}.
	\end{equation*}
%	\begin{enumerate}[label={\alph*)},leftmargin=1.4em]
%		\item Under event-level $\varepsilon$-differential privacy constraint, 
%		\begin{equation*}
%			\!\!\!\!\!\!\!\!\!\mE\left[ \left(\hat{\theta}(z) - \theta\right)\left(\hat{\theta}(z) - \theta\right)^\top \middle| \theta \right] \geq \Sigma_{\text{\rm CR}} + \frac{\left(\sum_{k=1}^K H_k^\top H_k\right)^{-1}}{K\varepsilon^2}; 
%		\end{equation*}
%		\item Under user-level $\varepsilon$-differential privacy constraint, 
%		\begin{equation*}
%			\!\!\!\!\!\!\!\!\!\mE\left[ \left(\hat{\theta}(z) - \theta\right)\left(\hat{\theta}(z) - \theta\right)^\top \middle| \theta \right] \geq \Sigma_{\text{\rm CR}} + \frac{\left(\sum_{k=1}^K H_k^\top H_k\right)^{-1}}{\varepsilon^2}. 
%		\end{equation*}
%	\end{enumerate}
\end{cor}
 
\begin{pf}
	The corollary follows from Proposition \ref{prop:DP2Fisher}, Corollary \ref{coro:PPCR} and $H^\top S H \leq \trace(S) H^\top H $,
%	\begin{align*}
%		H^\top S H \leq \trace(S) H^\top H 
%		= \trace(S) \sum_{k=1}^K H_k^\top H_k,
%	\end{align*}
	where $ H = $ $ \col\{H_1,\ldots,H_K\}$. 
\hfill $\qed$
\end{pf}

\begin{rem}
	From Corollary \ref{coro:DP}, we know that 
%	we know that the privacy-preserving CRLB theory provides a unified analytical framework for the impact of event-level and user-level differential privacy constraints on identification accuracy. Specifically, when the identification problem reduces to a mean estimation problem, $H_k = 1 $ for all $k$. In this case, the event-level 
	differential privacy introduces an additional MSE of order $O(1/K^2)$ to the mean estimation.
%	, while the user-level differential privacy introduces an additional MSE of order $O(1/K)$. 
%	This result is consistent with \cite{cai2021cost}. 
	Compared with similar results in \citet{cai2021cost,cummings2022mean,duchi2019lower,barnes2020fisher}, Corollary \ref{coro:DP} has the following advantages: 
	\begin{enumerate}[label={\roman*)},leftmargin=1.4em]
		\item Without privacy constraints, the privacy-preserving CRLB exactly reduces to the classical one. In contrast, the results in \citet{cai2021cost,cummings2022mean,duchi2019lower} only capture the convergence rate of $O(1/K)$, while the lower bound in \citet{barnes2020fisher} reduces to the trivial bound 0.
		
		\item $\{H_k\}$ can be any measurement matrix sequence ensuring system identifiability without requiring independent and identically distributed (i.i.d.) assumption.
		
%		is not required to be independent and identically distributed (i.i.d.). It can be any sequence ensuring system identifiability.
		
		\item For the additional MSE introduced by the differential privacy constraint, we provide an explicit lower bound without any unspecified constant factors. This allows for a precise, quantitative evaluation of the privacy-utility trade-off.
	\end{enumerate}
\end{rem}

\subsection{Privacy-Preserving Fisher Information Matrix and Its Recursive Calculation}

Similar to the inverse relationship between the classical CRLB and Fisher information matrix, we define the privacy-preserving Fisher information matrix as follows.
	
\begin{defn}
	Given the privacy constraint \( \mathcal{I}_z (y) \leq S \), the privacy-preserving Fisher information matrix 
%	for the measurement system \eqref{sys} 
	is defined as $\mathcal{PI}_{y|S}(\theta) = \Sigma_{\text{\rm PPCR}}^{-1}$.
%	\begin{align*}
%		\mathcal{PI}_{y|S}(\theta) = H^\top S^{\frac{1}{2}} \left( S^{\frac{1}{2}} \left(\mathcal{I}_{y}(H\theta)\right)^{-1} S^{\frac{1}{2}} + I_m \right)^{-1} S^{\frac{1}{2}} H.  
%	\end{align*}
\end{defn}

Notably, computing the privacy-preserving Fisher information for the multi-measurement system \eqref{sys:mult} requires inverting an $O(K)$-dimensional matrix, which incurs an $O(K^3)$ computational complexity. Therefore, this subsection focuses on recursive computation methods for both the privacy-preserving Fisher information and the corresponding privacy-preserving CRLB.

Denote
\begin{align}
	\bar{Y}_k =& \col\{y_1,\ldots,y_k\},\
	\bar{H}_k = \col\{H_1,\ldots,H_k\},\label{def:YH}\\
	\bar{Z}_k =& \col\{z_1,\ldots,z_k\},\
	\Sigma_{w,k} = \left(\mathcal{I}_{y_k}(H_k\theta)\right)^{-1},\nonumber\\
	&\quad \bar{\Sigma}_{w,k} = \diag\{\Sigma_{w,1},\ldots,\Sigma_{w,k}\}. \nonumber
\end{align}
%Then by the independence assumption on $\{w_k\}$, we have 
%\begin{align*}
%	\bar{\Sigma}_{w,k} = \diag\{\Sigma_{w,1},\ldots,\Sigma_{w,k}\}. 
%\end{align*}
The privacy constraint is formulated as $\mathcal{I}_{\bar{Z}_k}(\bar{Y}_k) \leq \bar{S}_k$, where  $\bar{S}_k = \begin{bmatrix}
	\bar{S}_{k-1} + \Delta \bar{S}_k & U_k \\
	U_k^\top & S_k
\end{bmatrix} $. By Lemma 1 of \citet{zamir1998Fisher}, 
$\mathcal{I}_{\bar{Z}_k}(\bar{Y}_k) - \mathcal{I}_{\bar{Z}_{k-1}}(\bar{Y}_k) = \mathcal{I}_{z_k}(\bar{Y}_k|\bar{Z}_{k-1}) $. Hence, $\begin{bmatrix}
\Delta \bar{S}_k & U_k \\
U_k^\top & S_k
\end{bmatrix}$ should be positive semi-definite, which implies $ \Delta \bar{S}_k \geq U_k S_k^+ U_k^\top $ and $U_k = U_k S_k^+ S_k$. Thus by minimizing $\Delta \bar{S}_k$, we obtain the  recursive privacy constraint
\begin{align}\label{eq:barS}
	\bar{S}_k = \begin{bmatrix}
		\bar{S}_{k-1} + U_k S_k^+ U_k^\top & U_k \\
		U_k^\top & S_k
	\end{bmatrix}.
\end{align}

Algorithm \ref{algo:recursive_PI} provides the recursive calculation of privacy-preserving Fisher information matrix. For notational convenience, denote $\mathcal{PI}_k = \mathcal{PI}_{\bar{Y}_k|\bar{S}_k}(\theta)$. 

\begin{algorithm}
	\caption{Recursive Computation of Privacy-Preserving Fisher Information Matrix}
	\label{algo:recursive_PI}
	\begin{algorithmic}
		\State \textbf{Input:}
		Sequences of measurement matrices $\{H_k\}$, inverse noise Fisher information matrices  $\{\Sigma_{w,k}\}$, and privacy levels \( \{\bar{S}_k\} \) with \eqref{eq:barS}. 
		\State \textbf{Output:}
		Sequence of Privacy-Preserving Fisher Information Matrices $\{\mathcal{PI}_k\}$.
		\State \textbf{Initialization:}  Set
		\vspace{-1em}
		\begin{align*}
			&B_1 = (I + S_1 \Sigma_{w,1})^{-1},\quad
			S_{\Delta,1} = (S_1 + \Sigma_{w,1}^{-1})^{-1},\\
			&\mathcal{PI}_1 = H_1^\top B_1 S_1 H_1. 
		\end{align*}
		\vspace{-1.5em}
		\For{$k = 2,3,\ldots$}
		\State Compute
		\vspace{-1em}
		\begin{align*}
			&\Phi_k = \begin{bmatrix} U_k S_k^+ \\ I \end{bmatrix},\quad
			\Gamma_k = \begin{bmatrix}
				B_{k-1} & 0 \\ 0 & I
			\end{bmatrix} \Phi_k,\\
			&S_{\Sigma,k} = \begin{bmatrix}
				S_{\Delta,k-1} & 0 \\ 0 & \Sigma_{w,k}
			\end{bmatrix},\quad
			\Psi_k = S_{\Sigma,k} \Phi_k \\
			&\Lambda_k = S_k \left(I+\Phi_k^\top S_{\Sigma,k} \Phi_k S_k \right)^{-1}, \\
			&S_{\Delta,k} = S_{\Sigma,k} - \Psi_k \Lambda_k \Psi_k^\top,\quad
			G_k = \bar{H}_k^\top \Gamma_k,\\
			&B_k = \begin{bmatrix}
				B_{k-1} & 0 \\ 0 & I
			\end{bmatrix} - \Gamma_k \Lambda_k \Psi_k^\top. 
		\end{align*}
		\vspace{-1em}
		\State Update $\mathcal{PI}_k = \mathcal{PI}_{k-1} + G_k \Lambda_k G_k^\top$.
%		\begin{align*}
%			\mathcal{PI}_k &= \mathcal{PI}_{k-1} + G_k \Lambda_k G_k^\top.
%		\end{align*}
%		\vspace{-1em}
		\EndFor
	\end{algorithmic}
\end{algorithm}

\begin{thm}\label{thm:PPF_recur}
	Suppose the condition of Corollary \ref{coro:DP} holds, and the privacy constraint is given by $\mathcal{I}_{\bar{Z}_k}(\bar{Y}_k) \leq \bar{S}_k$ for all $k$, where $\bar{S}_k$ admits the recursive structure in \eqref{eq:barS}. Then, the privacy-preserving Fisher information matrix $\mathcal{PI}_k$ can be recursively computed by Algorithm \ref{algo:recursive_PI}.
\end{thm}

The proof is provided in Appendix \ref{appen:proof}.

\begin{rem}
%	Algorithm \ref{algo:recursive_PI} decreases the single-step computational complexity of direct calculation from $O(k^3)$ to $O(k^2)$. This cost stems primarily from matrix multiplications involving $U_k$. Since the size of the new data $U_k$ at step $k$ scales as $O(k)$, an $O(1)$ complexity is generally unattainable. However, specific structures of $U_k$ allow further simplification. Specifically, w
	When $U_k = 0$, Algorithm \ref{algo:recursive_PI} can be simplified as $\mathcal{PI}_k = \mathcal{PI}_{k-1} + \mathcal{PI}_{y_k|S_k}(\theta).$ This additive property of the privacy-preserving Fisher information yields an $O(1)$ complexity.
\end{rem}

Then, we can get the recursive calculation of privacy-preserving CRLB $\Sigma_{\text{PPCR},k}$. 

\begin{algorithm}
	\caption{Recursive Computation of Privacy-Preserving CRLB}
	\label{algo:recursive_PPCR}
	\begin{algorithmic}
		\State \textbf{Input:}
		Sequences of privacy levels \( \{\bar{S}_k\} \) with \eqref{eq:barS}, and \( \{S_{\Sigma, k}, G_k,\Phi_k\} \)  from Algorithm \ref{algo:recursive_PI}.
		\State \textbf{Output:}
		Sequence of privacy-preserving CRLB $\{\Sigma_{\text{PPCR},k}\}$.
		\State \textbf{Initialization:}
		Set \( \Sigma_{\text{PPCR},k_0} = \mathcal{PI}_{k_0}^{-1} \). 
		\For{$k = k_0+1,k_0+2,\ldots$}
		\State Compute
		\vspace{-1em}
		\begin{align*}
			\Xi_k = I + \Phi_k^\top S_{\Sigma,k} \Phi_k S_k + G_k^\top \Sigma_{\text{PPCR},k-1} G_k S_k.
 		\end{align*}
 		\vspace{-1.5em}
 		\State Update
 		\vspace{-1em}
 		\begin{align*}
 			\Sigma_{\text{PPCR},k} = & \Sigma_{\text{PPCR},k-1} \\
 			& - \Sigma_{\text{PPCR},k-1} G_k S_k \Xi_k^{-1} G_k^\top \Sigma_{\text{PPCR},k-1}.
 		\end{align*}
 	 	\vspace{-2em}
		\EndFor
	\end{algorithmic}
\end{algorithm}

\begin{cor}\label{coro:PPCR_recur}
	Under the condition of Theorem \ref{thm:PPF_recur}, the privacy-preserving CRLB $\Sigma_{\text{PPCR},k}$ can be recursively computed by Algorithm \ref{algo:recursive_PPCR}. 
\end{cor}

\begin{pf}
	The corollary can be proved by Theorem \ref{thm:PPF_recur} and Theorem 1.17 of \citet{guo2002time}. 
\hfill $\qed$
\end{pf}

\subsection{Attainability and Rate Attainability}

This subsection will analyze the attainability of the privacy-preserving CRLB. 
%of the privacy-preserving CRLB. It is well known that, even without considering privacy, the classical CRLB is generally attainable only when the measurement noise is Gaussian \citep{lehmann1998point}. Hence for \textit{strict} attainability analysis, we restrict our analysis to the scenario where the measurement noises follow Gaussian distributions. 
Firstly, we consider the Gaussian measurement noise case. For the single-measurement system \eqref{sys} with $w\sim\mathcal{N}(\mu_w,\Sigma_w)$, one can prove that the following algorithm can attain the privacy-preserving CRLB.
\begin{align}
	&z = M(y,d) = S \left(y - \mu_w \right) + d,\ d \sim \mathcal{N}(0,S),\label{optM} \\
	&\hat{\theta}(z) = \Sigma_{\text{\rm PPCR}} H^\top  \left(  S \Sigma_w + \! I_m  \right)^{-1} z. \label{optEst}
\end{align}
For the multi-measurement system \eqref{sys:mult}, we will prove that the privacy-preserving RLS algorithm, i.e. Algorithm \ref{algo:PP_RLS}, can always attain the privacy-preserving CRLB. Without loss of generality, assume that $w_k$ is unbiased. 

\begin{algorithm}
	\caption{Privacy-Preserving RLS Algorithm}
	\label{algo:PP_RLS}
	\begin{algorithmic}
		\State \textbf{Input:}
		Sequences of measurements $\{y_k\}$, measurement matrices $\{H_k\}$, privacy levels \( \{\bar{S}_k\} \) with \eqref{eq:barS}, $\{\Phi_k, G_k,\Psi_k,\Lambda_k \}$ and $B_{k_0}$ from Algorithm \ref{algo:recursive_PI} and $\{\Sigma_{\text{PPCR},k}\}$ from Algorithm \ref{algo:recursive_PPCR}. 
		\State \textbf{Output:}
		Preserved output sequence $\{z_k\}$, and estimate sequence $ \{\hat{\theta}_{k}\} $.
		\State \textbf{Initialization:} Set $\zeta_0 = \varnothing$, $\bar Y_0=\varnothing$. 
		\For{$k = 1,2,\ldots$}
		\State \textbf{Phase 1: Sequential Gaussian Mechanism} 
		\State Generate privacy noise $d_k \sim \mathcal{N}(0, S_k)$ that is
		\State independent of $\bar{Y}_k$ and $d_1,\ldots,d_{k-1}$, and then
		\vspace{-1em}
		\begin{align*}
			z_k = U_k^\top \bar{Y}_{k-1} + S_k y_k + d_k.
		\end{align*}
		\vspace{-1.5em}
		\State \textbf{Phase 2: RLS Identification}
		\State Update $\zeta_k = \begin{bmatrix}
			\zeta_{k-1} \\ 0
		\end{bmatrix} + \Phi_k z_k$,
%		\begin{align*}
%			\zeta_k = & \begin{bmatrix}
%				\zeta_{k-1} \\ 0
%			\end{bmatrix} + \Phi_k z_k, 
%%			\\
%%			\hat{\theta}_k = & \hat{\theta}_{k-1} + \Sigma_{\text{PPCR},k} G_k \left( z_k - \Lambda_k \Psi_k^\top \zeta_k - \Lambda_k G_k^\top \hat{\theta}_{k-1}\right).
%		\end{align*}
		\If {$k=k_0$} 
		\( \hat\theta_{k_0} = \Sigma_{\mathrm{PPCR},k_0} \bar H_{k_0}^{\top}B_{k_0}\zeta_{k_0}. \)
%		\vspace{-1em} 
		\ElsIf{$k>k_0$} 
		\vspace{-1.5em}
		\State \[\!\!\!\!\! \hat\theta_k = \hat\theta_{k-1} + \Sigma_{\mathrm{PPCR},k}G_k \left( z_k-\Lambda_k\Psi_k^\top\zeta_k -\Lambda_kG_k^\top\hat\theta_{k-1} \right). \] 
		\vspace{-1.5em}
		\EndIf
		\EndFor
	\end{algorithmic}
\end{algorithm}

\begin{thm}\label{thm:opt}
	Under the condition of Theorem \ref{thm:PPF_recur} and Gaussian measurement noise assumption, Algorithm \ref{algo:PP_RLS} follows $\mathcal{I}_{\bar{Z}_k}(\bar{Y}_k) = \bar{S}_k$ for all $k \geq k_0$, and attains the privacy-preserving CRLB. 
\end{thm}

The proof is provided in Appendix \ref{appen:proof}.

Under non-Gaussian measurement noises, even without considering privacy, the classical CRLB is generally attainable only when the measurement noise is Gaussian \citep{lehmann1998point}. Thus, privacy-preserving CRLB is likewise unattainable. But, we can still show that the convergence rate of the gap between the privacy-preserving CRLB and the classical one, as characterized in Corollary \ref{coro:DP}, is attainable.

To avoid the complexity caused by the increasing dimension of $\bar{S}_k$, we relax the privacy constraint to $\trace(\mathcal{I}_{z_K}(\bar{Y}_K)) \leq s_K$ based on Proposition \ref{prop:DP2Fisher}. Under such a privacy constraint, we design the ML-based algorithm 
\begin{align}
	\hat{\theta}_{\text{ML}}(\bar{Y}_K) =& \arg\max_{\hat{\theta}} \sum_{k=1}^{K} l(y_k - H_k \hat{\theta}) \nonumber\\
	\hat{\theta}(z_K) =& z_K = \hat{\theta}_{\text{ML}}(\bar{Y}_K) + d_K, \label{algo:mult} \\ 
	d_K \sim& \mathcal{N}\left(0,\frac{n \mE (l^{\prime\prime}(w))^2}{s_K (\mE l^{\prime\prime}(w))^2} \left( \sum_{k=1}^K H_k^\top H_k\right)^{-1} \right), \nonumber
\end{align}
where $l(\cdot) = \ln f_w(\cdot) $.

\begin{assum}\label{assum:mult}
	$\{H_k\}$ and $\{w_k\}$ sequences satisfy
	\begin{enumerate}[leftmargin = 1.5em,label={\rm \roman*)}]
		\item $\sup_k \Abs{H_k} < \infty $ and $\lim_{K\to\infty}\frac{1}{K} \sum_{k=1}^K H_k^\top H_k \to \Sigma_H$
%		\begin{equation}%\label{condi:H_infty}
%			\lim_{K\to\infty}\frac{1}{K} \sum_{k=1}^K H_k^\top H_k \to \Sigma_H
%		\end{equation}
		for some positive definite $\Sigma_H$;
		\item $\{w_k\}$ is i.i.d., and $\mathcal{I}_{w_k}(\theta) = 0$;
		\item The density function $f_w(\cdot)$ is four times continuously differentiable, and $\max_{k = 1,\ldots, 4} \mathbb{E} \left[ \left| l^{(k)}(w) \right|^8 \right] < \infty $.
%		\item $ l(w) $ has a unique local maximum. 
	\end{enumerate}
\end{assum}

\begin{rem}
	Assumption \ref{assum:mult} iii) is used to ensure the asymptotic efficiency of ML algorithm. Therefore, noise distributions such as Laplacian one, which do not satisfy Assumption \ref{assum:mult} iii) but can still ensure the asymptotic efficiency of the ML algorithm, are also within the scope of the following theorem.
\end{rem}

\begin{thm}\label{thm:rate_attainablility}
	Under Assumption \ref{assum:mult}, $z_K$ in the algorithm \eqref{algo:mult} satisfies $ \lim_{K\to \infty} \trace(\mathcal{I}_{z_K}(\bar{Y}_K))/s_K = 1 $ almost surely. %, where $\mathcal{I}_{z_K}(\bar{Y}_K) = \bar{S}_K$.
	Therefore, 
	\begin{align}\label{concl:ML}
		& \mE\left[ \left(\hat{\theta}(z_K) - \theta\right)\left(\hat{\theta}(z_K) - \theta\right)^\top \middle| \theta \right] \nonumber\\
		= & \mE\left[ \left(\hat{\theta}_{\text{ML}}(\bar{Y}_K)- \theta\right)\left(\hat{\theta}_{\text{ML}}(\bar{Y}_K) - \theta\right)^\top \middle| \theta \right] \nonumber\\
		& + \frac{n \mE (l^{\prime\prime}(w))^2}{s_K (\mE l^{\prime\prime}(w))^2} \left( \sum_{k=1}^K H_k^\top H_k\right)^{-1} \nonumber\\
		= & \Sigma_{\text{\rm CR},K} + O\left(1/K^2\right) + \frac{n \mE (l^{\prime\prime}(w))^2}{s_K (\mE l^{\prime\prime}(w))^2} \left( \sum_{k=1}^K H_k^\top H_k\right)^{-1}\!\!, 
	\end{align}
	where $\Sigma_{\text{\rm CR},K}=(\mathcal{I}_{\bar{Y}_K}(\theta))^{-1} $ is the classical CRLB. 
\end{thm}

The proof is provided in Appendix \ref{appen:proof}.

\begin{rem}
	When $s_K = O(K)$, the algorithm \eqref{algo:mult} attains an MSE that is within a gap of $O(1/K^2)$ from the classical CRLB. According to Proposition \ref{prop:DP2Fisher}, such a result matches that in \citet{cai2021cost}.
\end{rem}

\section{Experimental Results}\label{sec:simu}

%This section employs numerical experiments to validate the privacy-preserving CRLB theory, the privacy-preserving efficiency of Algorithm \ref{algo:average}, and the privacy-preserving asymptotic efficiency of Algorithm \ref{algo:distributed}.

\subsection{Validation of Privacy-Preserving CRLB}

This sub-section employs a numerical experiment to validate the privacy-preserving CRLB theory.

In measurement system \eqref{sys}, set 
$$ \theta = \begin{bmatrix}
	0.63 & 0.81 & -0.75 & 0.83 & 0.26
\end{bmatrix}^\top, $$ 
and $H\in \mathbb{R}^{10 \times 5}$ is randomly selected from $ [-1,1]^{10\times 5} $. The measurement noise $w \sim \mathcal{N}(0,(0.2)^2 I_{10})$. The privacy constraint is \( \mathcal{I}_z (y) \leq S \) with \( S = s I_{10} \), where \( s \) ranges over \([0.1,10]\) in increments of \( 0.1 \). 

For comparison, we consider the following stochastic obfuscation mechanisms: 

\begin{itemize}[leftmargin = 1em]
	\item Gaussian mechanism \eqref{optM} with identification algorithm \eqref{optEst}.
	\item Laplacian mechanisms, including data perturbation \citep{wang2024sample} and output perturbation types\footnote{The data perturbation type directly perturbs the sensitive data $y$ by using privacy noise $d$, while the output perturbation type perturbs the original estimate \citep{wang2024sample,turgay2023perturbation}.} \citep{dwork2006calibrating}. For data perturbation type, we select the ML algorithm to ensure the optimality of identification accuracy. For output perturbation type, the output $z$ is directly considered as the final estimate as \citet{dwork2006calibrating,cummings2022mean} do. 
	\item Cauchy mechanism of data-perturbation type  \citep{ito2021privacy} with the ML identification algorithm.
	\item Trigonometry-based mechanism \eqref{mech:trigonometry} in Appendix \ref{appen:examples} of output-perturbation type with output directly as the estimate \citep{farokhi2019ensuring}.
	\item Multiplicative noise mechanism of data-perturbation type \citep{ye2024privacy} with the ML identification algorithm. The inverse of privacy noises follows unbiased Gaussian distribution. The stochastic truncation technique proposed in Remark \ref{remark:truncation} is used to ensure the privacy constraint.  
	\item Probabilistic quantization mechanism of output-perturbation type with output directly as the estimate \citep{wang2023quantization}. A stochastic truncation technique similar to that in Remark \ref{remark:truncation} is used to ensure the privacy constraint.
\end{itemize}

\begin{figure}[!htbp]
	\centering
	\includegraphics[width=0.9\linewidth]{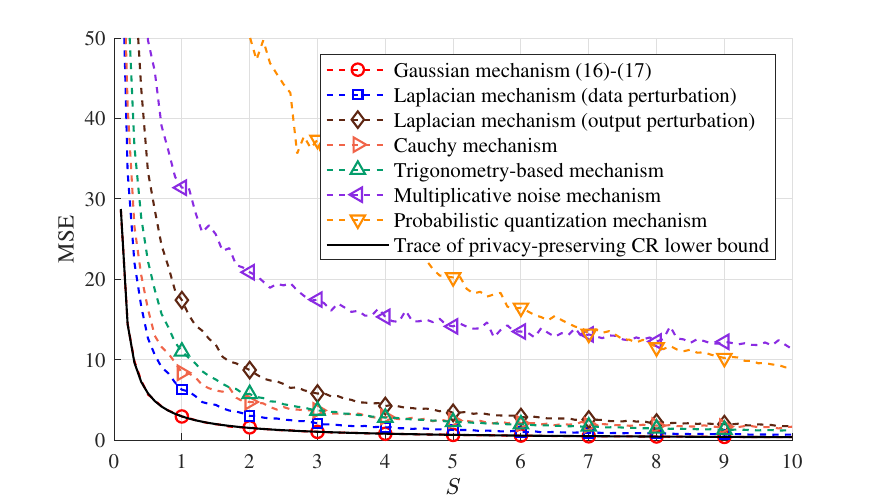}
	\caption{MSEs under different mechanisms}
	\label{fig:ppcr}
\end{figure}

2000 repeated experiments are conducted for each algorithm at each privacy level $S$ to obtain the MSE. As shown in Fig. \ref{fig:ppcr}, the Gaussian mechanism \eqref{optM} with the identification algorithm \eqref{optEst} attains the privacy-preserving CRLB, while other mechanisms do not, which demonstrates the optimality of the Gaussian mechanism.  

\subsection{Experiment on Recursive Identification}

In this subsection, Algorithm \ref{algo:PP_RLS} is applied to recursive identification for the serum triglyceride level analysis\footnote{Data come from the National Health and Nutrition Examination Survey.}. Specifically, \(y_k\) is the Box-Cox transformed serum triglyceride level, whose distribution is approximately Gaussian, and
\(H_k\) represents the features of participant \(k\), including an intercept term, age, body mass index, sex, and income-to-poverty ratio. 
The goal is to learn a simple linear relation between these participant features and the transformed triglyceride level from real data. 

\begin{figure}[htbp]
	\centering
	\includegraphics[width=0.9\linewidth]{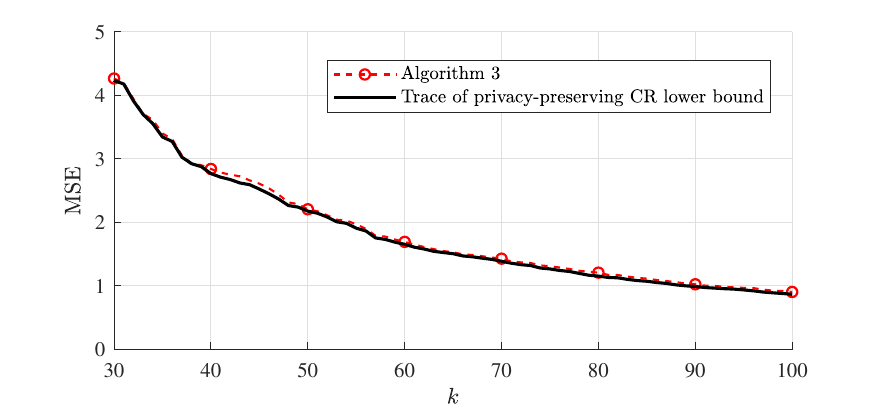}
	\caption{MSEs for Algorithm \ref{algo:PP_RLS}}
	\label{fig:pprls-non-syms}
\end{figure}

The 2704 participants are randomly split into an experimental set of 2000 samples and a reference set containing the remaining samples. The estimate obtained from the reference set is treated as the ground-truth parameter, and its residual variance is used as the noise variance. The experimental set is divided into 20 real-data blocks, each containing 100 samples. Algorithm \ref{algo:PP_RLS} is applied to each block with 20 independent repetitions to average out the randomness caused by privacy noise. 
The cumulative privacy constraint \(\bar{S}_k\) is generated recursively by \eqref{eq:barS}, where each \(S_k\) is uniformly drawn from $[0.2,2]$, \(U_k=0.1\sqrt{S_k}\xi_k\), and \(\xi_k\) follows the standard Gaussian distribution. As shown in Fig. \ref{fig:pprls-non-syms}, Algorithm \ref{algo:PP_RLS} closely matches the privacy-preserving CRLB. 

\begin{figure}[htbp]
	\centering
	\includegraphics[width=0.9\linewidth]{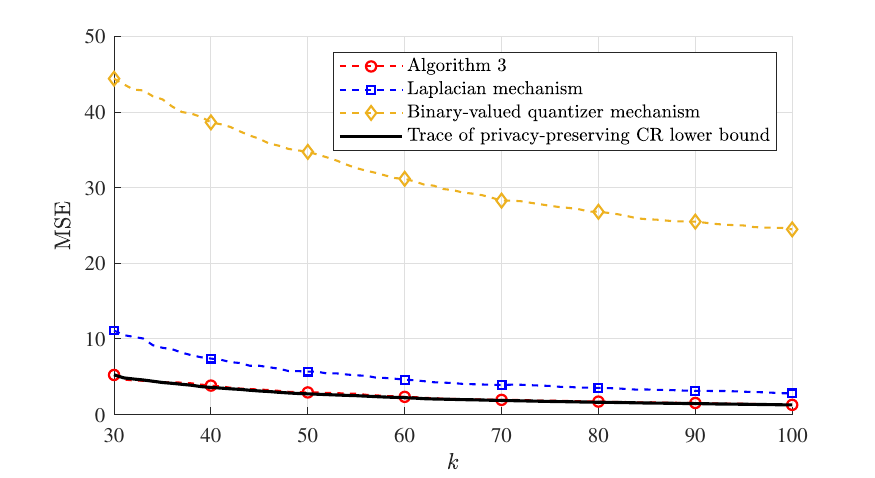}
	\caption{MSEs for recursive identification algorithms}
	\label{fig:pprls-syms}
\end{figure}

To compare with existing mechanisms, we consider the case where
$\bar{S}_k=I$. The compared mechanisms include Laplacian mechanism \citep{liu2018differentially} with the RLS identification algorithm, and the
binary-valued quantization mechanism based on Gaussian
privacy noises \citep{ke2025privacy+quantization} with an asymptotically optimal quasi-Newton type identification algorithm \citep{wang2024asymptotically}. Fig. \ref{fig:pprls-syms} shows that Algorithm \ref{algo:PP_RLS} performs better than these two mechanisms. 

\subsection{Experiment Under Laplacian Measurements}

\begin{figure}[H]
	\centering
	\includegraphics[width=0.9\linewidth]{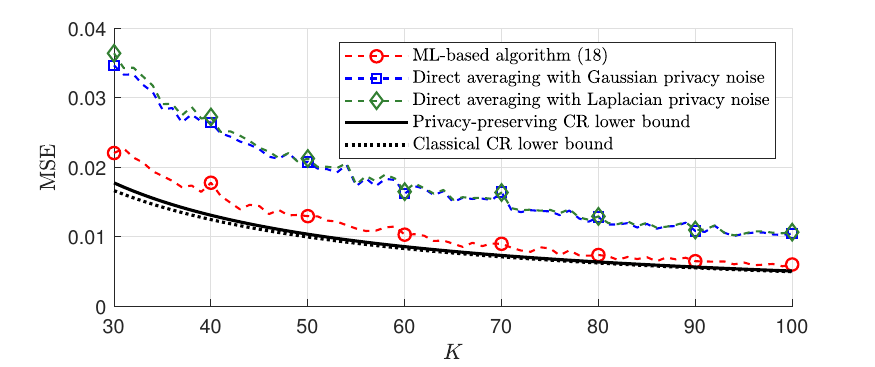}
	\caption{MSEs for ML-based algorithm and direct averaging ones}
	\label{fig:ml}
\end{figure}

In this subsection, we validate the ML-based privacy-preserving algorithm \eqref{algo:mult} under non-Gaussian measurement noises. 
To compare the algorithm \eqref{algo:mult} with those in \cite{cai2021cost,cummings2022mean}, 
we focus on the mean estimation problem. Specifically, $y_k=\theta+w_k$, where $\theta = 0.53$ and $w_k$ follows an unbiased Laplacian distribution with variance 1. We consider the privacy constraint $ \lim_{K\to \infty} \trace(\mathcal{I}_{z_K}(\bar{Y}_K))/K = 1 $, and compare the algorithm \eqref{algo:mult} with the direct averaging methods with Gaussian and Laplacian privacy noises \citep{cai2021cost}. Fig. \ref{fig:ml} shows that the algorithm \eqref{algo:mult} is close to the privacy-preserving CRLB, while the direct averaging methods have larger MSEs due to the efficiency loss caused by non-Gaussian measurement noises.

\section{Conclusion}\label{sec:concl}

This paper established the privacy-preserving CRLB theory for system identification.
Recursive computation formulas were derived for the proposed lower bound, and its attainability was analyzed.
This theory characterizes the foundational trade-off between identification accuracy and privacy preservation, independent of any specific privacy mechanism or identification algorithm.
Specifically, under the privacy constraint $\mathcal{I}_z(y)\leq S$, the minimum attainable MSE is at least $(H^\top S H)^{-1}$ larger than the classical CRLB. 
By further building a quantitative relationship between differential privacy and Fisher information, the proposed results can also be extended to the differential privacy framework.

The privacy-preserving CRLB provides a fundamental optimality benchmark for privacy-preserving system identification. With this benchmark, Theorem \ref{thm:opt} provides an initial result for privacy-preserving optimal identification, which can be further developed toward related problems such as privacy-preserving Kalman filtering.
%a theory of privacy-preserving optimal identification can be developed. 
Moreover, general privacy-preserving identification algorithms can be evaluated by comparing their MSEs with this lower bound. This comparison separates the intrinsic accuracy loss caused by privacy preservation from the additional loss caused by non-optimal algorithm design. 
%Another interesting topic is to investigate the dual problem of the privacy-preserving CRLB, i.e., the optimal allocation of privacy budgets under a prescribed identification accuracy.

%This paper establishes an identifiability criterion under privacy constraint and derives the privacy-preserving CRLB under identifiable conditions. This bound defines the minimum mean squared identification error achievable by any unbiased estimator when the privacy constraint is described using Fisher information matrix. Notably, the proposed privacy-preserving CRLB is precise and attainable. Gaussian mechanism with its corresponding least squares algorithm is shown to achieve this bound. Therefore, the privacy-preserving Fisher information matrix exists with explicit expression. 
%The privacy-preserving CRLB theory is further extended to the multi-sensor multi-measurement system.
%Specifically, an additivity principle of privacy-preserving Fisher information matrices is established across both spatial and temporal scales. Based on this principle, two distributed identification algorithms are proposed, including a privacy-preserving distributed offline algorithm that converges to an efficient estimate and a privacy-preserving distributed online algorithm that achieves privacy-preserving asymptotic efficiency. Future works can focus on the extension to nonlinear systems. 

\appendix

\section{Admissible Privacy Mechanisms}\label{appen:examples}

This appendix presents several examples to illustrate that Definition \ref{def:admissible} encompasses a broad class of admissible mechanisms.

\begin{exmp}\label{ex:affine}
	\noindent{\normalfont\bfseries (Affine transformation mechanisms).}\ The affine transformation mechanisms are
	\begin{equation}
		z = M(y,d) = A y + b + B d
	\end{equation}
	where $A$ and $B$ are given matrices, $b $ is a given vector,
%	$A,\ B $ and $b $ are given matrices or vector, 
	and $d$ denotes privacy noise. The affine transformation mechanism is most common in existing works on privacy preservation for real-valued sensitive data. 
	Based on different privacy noises $d$, this mechanism can be categorized into:
	\begin{itemize}[leftmargin=1.4em]
		\item \textbf{Laplacian mechanisms} \citep{dwork2009what,chen2024locally}: 
		%		Affine transformation mechanism where 
		$d$ is Laplacian distributed.
		\item \textbf{Gaussian mechanisms} \citep{asoodeh2018estimation,dong2022gaussian}:
		%		Affine transformation mechanism where 
		$d$ is Gaussian distributed.
		\item \textbf{Bounded noise mechanisms} \citep{geng2016optimal,farokhi2019ensuring,dagan2022bounded}: $d$ has bounded support, e.g.,
		%		follows noise distributions with bounded supports, 
		%		such as 
		the noises with 
		%		the uniform distribution \cite{geng2016optimal,farokhi2019ensuring,dagan2022bounded} investigate optimal stochastic obfuscation mechanisms under different bounded privacy noises, including the uniformly distributed noise and noises $d$ that follow 
		density functions
		\begin{equation}\label{mech:trigonometry}
			\!\!\!\!\!\!\!\! \left( \frac{2}{\bar{d} - \underline{d}} \right)^{\bar{n}} \prod_{i = 1}^{\bar{n}} \left(\cos \left( \frac{\pi}{\bar{d} - \underline{d}} \left( d_i - \frac{\bar{d} + \underline{d}}{2} \right) \right)\right)^2 \mathbb{I}_{\left\{ d \in \left[\underline{d},\bar{d}\right]^{\bar{n}} \right\}} 
		\end{equation}
		and
		\begin{equation*}
			\frac{1}{C} \exp \left( - \frac{1}{\left(1-d^2\right)^p} \right) \mathbb{I}_{\left\{ d \in \left(-1,1\right) \right\}} 
		\end{equation*}
		where $d_i, i = 1,\ldots,\bar{n}$ denote components of $d$, and $p,C$ are given constants.
		\item \textbf{Other mechanisms}: $d$ follows other distributions such as heavy-tailed ones \citep{ito2021privacy} or staircase-shaped one \citep{geng2016optimal}.
		%		For instance, \cite{ito2021privacy} considers affine transformation mechanisms with heavy-tailed distribution noise $d$, while \cite{geng2016optimal} designs affine transformation mechanisms with staircase-shaped density functions beyond Laplacian mechanisms.
	\end{itemize}
\end{exmp}

\begin{exmp}
	\noindent{\normalfont\bfseries (Multiplicative and mixed noise mechanisms).}\ Beyond affine transformation mechanisms, multiplicative and mixed noise mechanisms are also admissible.
	\begin{itemize}[leftmargin=1.4em]
		\item \textbf{Multiplicative noise mechanisms:}  $z = M(y,D) = D (A y + b)$, where $A$ and $b$ are defined as in Example \ref{ex:affine}, and $D$ represents privacy noise on $\mathbb{R}^{\bar{m} \times \bar{m}}$. Specifically, \citet{brackenbury2019protecting} design mechanisms with twin uniformly distributed privacy noise, whereas \citet{ye2024privacy} investigate mechanisms where the inverse of privacy noises follows unbiased Laplacian and Gaussian distributions. Additionally, the random-rotation-perturbation method \citep{chen2005privacy} and dynamics-based method \citep{wang2022decentralized} also fall under multiplicative noise mechanisms. 
		\item \textbf{Mixed noise mechanisms:} \( z = M(y,d,D) = D (A y + b) + B d \), where \( A, B, b, D, d \) are defined as in Example \ref{ex:affine} and the above discussion about multiplicative noise mechanisms. Combining affine transformation and multiplicative noise, works such as \citet{ke2023differentiated,chen2011geometric} investigate algorithms that fit into this category. In \citet{ke2023differentiated}, privacy weights serve as multiplicative noises while Gaussian privacy noises are additive in the affine transformation component. Furthermore, the geometric data perturbation framework proposed in \citet{chen2011geometric} integrates multiplicative noise through random geometric transformations and additive noise for distance perturbation.
	\end{itemize}
\end{exmp}

\begin{rem}\label{remark:truncation}
	$\mathcal{I}_z(y)$ for multiplicative noise mechanisms diverges at $Ay+b=0$. To address this issue, 
	we can preprocess $Ay+b$ using a probabilistic quantizer \citep{liu2026design} that maps each element of $Ay+b$ lying in $[-\epsilon,\epsilon]$ to either $-\epsilon$ or $\epsilon$, ensuring uniformly bounded Fisher information for the modified mechanism, where $\epsilon>0$ is a small constant. Such a stochastic truncation technique can also be adopted for other mechanisms suffering from unbounded Fisher information. 
\end{rem}

\begin{exmp}
	\noindent{\normalfont\bfseries (Quantizer-based mechanisms).}\ Quan -tizers can be utilized to preserve privacy while simultaneously enhancing communication efficiency. The quantizer-based mechanism can be expressed as
	$$z = M(y,d) = \mathcal{Q}(Ay + d),$$
	where $\mathcal{Q}(\cdot)$ denotes the quantizer, and $A$ and $d$ are defined as in Example \ref{ex:affine}.
	There are several different quantizer-based mechanisms.
	\begin{itemize}[leftmargin=1.4em]
		\item \textbf{Probabilistic quantization  mechanisms} \citep{liu2026design,wang2023quantization}: $\mathcal{Q}(\cdot)$ is a uniform quantizer and $d$ is an unbiased uniformly distributed noise. The support of this uniform noise precisely matches the quantization step size. 
		\item \textbf{Laplacian quantization mechanism} \citep{lang2023federated}: $\mathcal{Q}(\cdot)$ is a uniform quantizer and $d$ follows a Laplacian distribution. 
		%		The Laplacian noise must be sufficiently large so that it can be decomposed into the sum of a uniformly distributed noise whose support matches the quantization step size and an additional independent unbiased noise component.
		\item \textbf{Binary-valued quantization mechanism} \citep{ke2025privacy+quantization}: $\mathcal{Q}(\cdot)$ is a binary-valued quantizer, while $d$ can be Gaussian, Laplacian, or Cauchy.%, whose density function exhibits a uniform positive lower bound on any compact set and possesses bounded Fisher information. 
	\end{itemize}
\end{exmp}

\section{Supplementary Proofs}\label{appen:proof}

%\subsection{Proof of Corollary \ref{coro:PPCR}}

\textbf{Proof of Corollary \ref{coro:PPCR}.} Define $ S_\epsilon = (S^{\frac{1}{2}} + \epsilon I_m)^2 $ as in the proof of Theorem \ref{thm:crbound}. Then, we have 
\begin{align*}%\label{eq:PPCR_e}
	\Sigma_{\text{\rm PPCR}}^{-1} = \lim_{\epsilon \to 0} H^\top S_\epsilon^{\frac{1}{2}} \left( S_\epsilon^{\frac{1}{2}} \left(\mathcal{I}_{y}(H\theta)\right)^{-1} S_\epsilon^{\frac{1}{2}} + I_m \right)^{-1} S_\epsilon^{\frac{1}{2}} H.
\end{align*}
Since $S_\epsilon$ is invertible, by Theorem 1.17 of \citet{guo2002time}, %one can get
\begin{align*}
	& S_\epsilon^{\frac{1}{2}} \left( S_\epsilon^{\frac{1}{2}} \left(\mathcal{I}_{y}(H\theta)\right)^{-1} S_\epsilon^{\frac{1}{2}} + I_m \right)^{-1} S_\epsilon^{\frac{1}{2}} \\
	= & \left( \mathcal{I}_{y}(H\theta)^{-1} + S_\epsilon^{-1} \right)^{-1} \\
	= & \mathcal{I}_{y}(H\theta) - \mathcal{I}_{y}(H\theta) \left(\mathcal{I}_{y}(H\theta) + S_\epsilon \right)^{-1} \mathcal{I}_{y}(H\theta).  
\end{align*}
Letting $\epsilon \to 0$ and by Lemma 4 of \citet{zamir1998Fisher}, we have
\begin{align}\label{eq:PPCR-CR}
	\Sigma_{\text{\rm PPCR}}^{-1} 
%	\! = &  \mathcal{I}_{y}(\theta) \! - \! H^\top  \mathcal{I}_{y}(H\theta) \left(\mathcal{I}_{y}(H\theta) + S \right)^{-1} \mathcal{I}_{y}(H\theta) H \nonumber\\
	= & \Sigma_{\text{\rm CR}}^{-1} - H^\top  \mathcal{I}_{y}(H\theta) \left(\mathcal{I}_{y}(H\theta) + S \right)^{-1} \mathcal{I}_{y}(H\theta) H
\end{align} 

Note that $P^{-1} \geq  H \left(H^\top P H \right)^{-1} H^\top$ for $ P = \mathcal{I}_{y}(H\theta) + S $.
%\begin{align*}
%	& \left(\mathcal{I}_{y}(H\theta) + S \right)^{-1} - H \left(H^\top \mathcal{I}_{y}(H\theta) H + H^\top S H \right)^{-1} H^\top \\
%	= & \left(\mathcal{I}_{y}(H\theta) + S \right)^{-\frac{1}{2}} \left( I_m - \left(\mathcal{I}_{y}(H\theta) + S \right)^{\frac{1}{2}} H  \right. \\
%	& \left. \cdot \left(H^\top \left(\mathcal{I}_{y}(H\theta) + S\right) H \right)^{-1} H^\top \left(\mathcal{I}_{y}(H\theta) + S \right)^{\frac{1}{2}} \right) \\
%	&  \cdot \left(\mathcal{I}_{y}(H\theta) + S \right)^{-\frac{1}{2}} \\
%	= & \left(\mathcal{I}_{y}(H\theta) + S \right)^{-\frac{1}{2}} \left( I_m - \left(\mathcal{I}_{y}(H\theta) + S \right)^{\frac{1}{2}} H  \right. \\
%	&\left. \cdot \left(H^\top \left(\mathcal{I}_{y}(H\theta) + S\right) H \right)^{-1} H^\top \left(\mathcal{I}_{y}(H\theta) + S \right)^{\frac{1}{2}} \right)^2 \\ & \cdot \left(\mathcal{I}_{y}(H\theta) + S \right)^{-\frac{1}{2}}
%	\geq 0. 
%\end{align*}
Then, we have
\begin{align*}
	& H^\top  \mathcal{I}_{y}(H\theta) \left(\mathcal{I}_{y}(H\theta) + S \right)^{-1} \mathcal{I}_{y}(H\theta) H \\
	\geq & \mathcal{I}_{y}(\theta) \left(H^\top \mathcal{I}_{y}(H\theta) H + H^\top S H \right)^{-1} \mathcal{I}_{y}(\theta)  \\
	\geq & \Sigma_{\text{\rm CR}}^{-1} \left(\Sigma_{\text{\rm CR}}^{-1} + H^\top S H \right)^{-1} \Sigma_{\text{\rm CR}}^{-1}. 
\end{align*}
Substituting the above inequality to \eqref{eq:PPCR-CR}, by Theorem 1.17 of \citet{guo2002time}, one can get
\begin{align*}
	\Sigma_{\text{\rm PPCR}}^{-1} 
	\leq & \Sigma_{\text{\rm CR}}^{-1} - \Sigma_{\text{\rm CR}}^{-1} \left(\Sigma_{\text{\rm CR}}^{-1} + H^\top S H \right)^{-1} \Sigma_{\text{\rm CR}}^{-1} \\
	= & \left( \Sigma_{\text{\rm CR}} + \left(H^\top S H\right)^{-1} \right)^{-1},
\end{align*}
which implies the corollary. 	
\hfill $\qed$

\begin{lemx} \label{lemma:PPCR_S}
	For positive semi-definite  $S,\ \Sigma $ and $ L $ with $S = L L^\top $, $L \left( L^\top \Sigma L + I \right)^{-1} L^\top = (I + S \Sigma)^{-1} S $.
%	\begin{align*} 
%		L \left( L^\top \Sigma L + I \right)^{-1} L^\top = (I + S \Sigma)^{-1} S = S (I + \Sigma S)^{-1}\! .
%	\end{align*}
\end{lemx}

\begin{pf}
	Let $M \triangleq L \left( L^\top \Sigma L + I \right)^{-1} L^\top$. We first prove the left equality by evaluating the product $(I + S \Sigma) M$. 
	
	Note that $(I + S \Sigma) L = L + L L^\top \Sigma L
	= L \left( I + L^\top \Sigma L \right)$.
%	\begin{align*}
%		(I + S \Sigma) L &= L + L L^\top \Sigma L
%		= L \left( I + L^\top \Sigma L \right).
%	\end{align*}
	Then, we have
	\begin{align*}
		&(I + S \Sigma) M = \left[ (I + S \Sigma) L \right] \left( L^\top \Sigma L + I \right)^{-1} L^\top \\
		&= L \left( I + L^\top \Sigma L \right) \left( L^\top \Sigma L + I \right)^{-1} L^\top 
		= L I L^\top 
		= S.
	\end{align*}
	Since $S$ and $\Sigma$ are positive semi-definite, the eigenvalues of $L^\top \Sigma L$ are non-negative. For any $\lambda$, 
	
	\begin{align*}
		\det\left(\lambda I - L^\top \Sigma L \right) 
		= & \det \begin{bmatrix}
			I & L\\
			0 & \lambda I - L^\top \Sigma L
		\end{bmatrix} 
		= \det \begin{bmatrix}
			I & L\\
			L^\top \Sigma  & \lambda I
		\end{bmatrix} \\
		= & \det \begin{bmatrix}
			I - \frac{1}{\lambda} S \Sigma & L\\
			0 & \lambda I
		\end{bmatrix}
		= \det\left(\lambda I - S \Sigma \right),
	\end{align*}
	which guarantees that $I + S \Sigma$ is non-singular. Hence, $M = (I + S \Sigma)^{-1} S$.
%	\begin{align*}
%		M = (I + S \Sigma)^{-1} S.
%	\end{align*}
%	Furthermore, $M = M^\top = S (I + \Sigma S)^{-1}$.
%	\begin{align*}
%		M = M^\top = S (I + \Sigma S)^{-1}. 
%	\end{align*}
	This completes the proof.
	\hfill $\qed$
\end{pf}

\textbf{Proof of Theorem \ref{thm:PPF_recur}.}
Firstly, we will prove $S_{\Delta,k} = (\bar{S}_k + \bar{\Sigma}_{w,k}^{-1})^{-1} $. 

By \eqref{eq:barS}, 
\begin{align}\label{eq:recur_barS}
	\bar{S}_k = \begin{bmatrix}
		\bar{S}_{k-1} & 0 \\ 0 & 0
	\end{bmatrix} + \Phi_k S_k \Phi_k^\top.
\end{align}
Hence by Theorem 1.17 of \citet{guo2002time},
\begin{align*}
	S_{\Delta,k} 
	= & S_{\Sigma,k} - S_{\Sigma,k} \Phi_k S_k \left(I+\Phi_k^\top S_{\Sigma,k} \Phi_k S_k \right)^{-1} \Phi_k^\top S_{\Sigma,k} \\
	= & \left(S_{\Sigma,k}^{-1} + \Phi_k S_k \Phi_k^\top \right)^{-1}.
\end{align*}
%	where $S_{\Sigma,k} = \begin{bmatrix}
	%		S_{\Delta,k-1} & 0 \\ 0 & \Sigma_{w,k}
	%	\end{bmatrix}$.
Then when $S_{\Delta,k-1} = (\bar{S}_{k-1} + \bar{\Sigma}_{w,k-1}^{-1})^{-1}$, 
\begin{align*}
	& S_{\Delta,k}
	= \left(\begin{bmatrix}
		S_{\Delta,k-1}^{-1} & 0 \\ 0 & \Sigma_{w,k}^{-1}
	\end{bmatrix} + \Phi_k S_k \Phi_k^\top \right)^{-1} \\
	= & \left(\bar{\Sigma}_k^{-1} + \begin{bmatrix}
		\bar{S}_{k-1} & 0 \\ 0 & 0
	\end{bmatrix} + \Phi_k S_k \Phi_k^\top \right)^{-1}
	= (\bar{S}_k + \bar{\Sigma}_{w,k}^{-1})^{-1}. 
\end{align*}

Secondly, we will prove $B_k = (I + \bar{S}_k \bar{\Sigma}_{w,k})^{-1} $ by induction.

Assume that $B_{k-1} = (I + \bar{S}_{k-1} \bar{\Sigma}_{w,k-1})^{-1}$. Then, $B_{k-1} = \bar{\Sigma}_{w,k-1}^{-1} S_{\Delta,k-1}$. Therefore,
\begin{align*}
	& (I + \bar{S}_k \bar{\Sigma}_{w,k})^{-1}
	= \bar{\Sigma}_{w,k}^{-1} S_{\Delta,k} \\
	= & \bar{\Sigma}_{w,k}^{-1} \left( \begin{bmatrix}
		S_{\Delta,k-1} & 0 \\ 0 & \Sigma_{w,k}
	\end{bmatrix} - \Psi_k \Lambda_k \Psi_k^\top \right) \\
	= & \begin{bmatrix}
		B_{k-1} & 0 \\ 0 & I
	\end{bmatrix} - \bar{\Sigma}_{w,k}^{-1}  \Psi_k \Lambda_k \Psi_k^\top \\
	= & \begin{bmatrix}
		B_{k-1} & 0 \\ 0 & I
	\end{bmatrix} - \begin{bmatrix}
		B_{k-1} & 0 \\ 0 & I
	\end{bmatrix} \Phi_k \Lambda_k \Psi_k^\top 
	%		\\
	%		= & \begin{bmatrix}
		%			B_{k-1} & 0 \\ 0 & I
		%		\end{bmatrix} - \Gamma_k \Lambda_k \Psi_k^\top
	= B_k. 
\end{align*}

Thirdly, we prove the theorem. By Lemma \ref{lemma:PPCR_S}, it suffices to prove that $\bar{X}_1 = B_1 S_1 $ and $\bar{X}_k = \begin{bmatrix}
	\bar{X}_{k-1} & 0 \\ 0 & 0 
\end{bmatrix} + \Gamma_k \Lambda_k \Gamma_k^\top$
%\begin{align*}
%	\bar{X}_k = \begin{bmatrix}
%		\bar{X}_{k-1} & 0 \\ 0 & 0 
%	\end{bmatrix} + \Gamma_k \Lambda_k \Gamma_k^\top
%\end{align*}
implies 
$ \bar{X}_k = B_k \bar{S}_k $. 

We again use induction. Assume that $\bar{X}_{k-1} = B_{k-1} \bar{S}_{k-1}$. Then, 
\begin{align*}
	& \bar{X}_{k-1} = (I + \bar{S}_{k-1} \bar{\Sigma}_{w,k-1})^{-1} \bar{S}_{k-1} \bar{\Sigma}_{w,k-1} \bar{\Sigma}_{w,k-1}^{-1} \\
	= & \bar{\Sigma}_{w,k-1}^{-1} - (I + \bar{S}_{k-1} \bar{\Sigma}_{w,k-1})^{-1} \bar{\Sigma}_{w,k-1}^{-1} \\
	= & \bar{\Sigma}_{w,k-1}^{-1} - \bar{\Sigma}_{w,k-1}^{-1} S_{\Delta,k-1} \bar{\Sigma}_{w,k-1}^{-1},
\end{align*}
which implies
\begin{align*}
	& \bar{\Sigma}_{w,k}^{-1} - \bar{\Sigma}_{w,k}^{-1} S_{\Sigma,k} \bar{\Sigma}_{w,k}^{-1} \\
	= & \begin{bmatrix}
		\bar{\Sigma}_{w,k-1}^{-1} - \! \bar{\Sigma}_{w,k-1}^{-1} S_{\Delta,k-1} \bar{\Sigma}_{w,k-1}^{-1} & \! 0 \\ 0 &
		\! \Sigma_{w,k}^{-1} - \! \Sigma_{w,k}^{-1} \Sigma_{w,k} \Sigma_{w,k}^{-1}
	\end{bmatrix} \\
	= & \begin{bmatrix}
		\bar{X}_{k-1} & 0 \\ 0 & 0 
	\end{bmatrix}. 
\end{align*}
Therefore,
\begin{align*}
	& B_k \bar{S}_k 
	= \bar{\Sigma}_{w,k}^{-1} - \bar{\Sigma}_{w,k}^{-1} S_{\Delta,k} \bar{\Sigma}_{w,k}^{-1}\\
	= & \bar{\Sigma}_{w,k}^{-1} - \bar{\Sigma}_{w,k}^{-1} \left(  S_{\Sigma,k} - \Psi_k \Lambda_k \Psi_k^\top \right) \bar{\Sigma}_{w,k}^{-1} \\
	= & \begin{bmatrix}
		\bar{X}_{k-1} & 0 \\ 0 & 0 
	\end{bmatrix} + \Gamma_k \Lambda_k \Gamma_k^\top
	= \bar{X}_k. 
\end{align*}
The theorem is thereby proved. 
\hfill $\qed$

\textbf{Proof of Theorem \ref{thm:opt}.}
Firstly, we will prove $\mathcal{I}_{\bar{Z}_k}(\bar{Y}_k) = \bar{S}_k$ by induction. 
Assume that $\mathcal{I}_{\bar{Z}_{k-1}}(\bar{Y}_{k-1}) = \bar{S}_{k-1}$. Then by the independence of $\{d_k\}$ and Lemma 1 of \citet{zamir1998Fisher}, 
\begin{align*}
	\mathcal{I}_{\bar{Z}_k}(\bar{Y}_k) = \mathcal{I}_{\bar{Z}_{k-1}}(\bar{Y}_{k}) +\mathcal{I}_{z_k}(\bar{Y}_{k})
	= \begin{bmatrix}
		\bar{S}_{k-1} & 0 \\ 0 & 0
	\end{bmatrix} + \mathcal{I}_{z_k}(\bar{Y}_{k}).
\end{align*}

Now, we compute $\mathcal{I}_{z_k}(\bar{Y}_{k})$.
Since, $\begin{bmatrix}
	\Delta \bar{S}_k & U_k \\
	U_k^\top & S_k
\end{bmatrix}$ is positive semi-definite, then 
\begin{align*}
	& \begin{bmatrix}
		I & 0 \\ 0 & I - S_k S_k^+ 
	\end{bmatrix}
	\begin{bmatrix}
		\Delta \bar{S}_k & U_k \\
		U_k^\top & S_k
	\end{bmatrix}
	\begin{bmatrix}
		I & 0 \\ 0 & I - S_k^+ S_k 
	\end{bmatrix} \\
	= & \begin{bmatrix}
		\Delta \bar{S}_k & U_k (I - S_k^+ S_k ) \\
		(I - S_k S_k^+) U_k^\top & 0
	\end{bmatrix} \geq 0,
\end{align*}
which implies that $ U_k = U_k S_k^+ S_k $. Additionally by $d_k \sim \mathcal{N}(0, S_k)$, the covariance of $ S_k S_k^+ d_k - d_k $ is strictly $0$, which means $ S_k S_k^+ d_k = d_k $. Hence, 
\begin{align*}
	S_k^+ S_k z_k = &
	S_k S_k^+ z_k = S_k S_k^+ U^\top \bar{Y}_{k-1} + S_k S_k^+ S_k y_k + d_k \\
	= & U_k^\top \bar{Y}_{k-1} + S_k y_k + d_k = z_k
\end{align*}
Since $S_k$ is positive semi-definite, there exists $\ell_k$ such that $S_k = \ell_k \ell_k^\top$ and $\ell_k^\top \ell_k$ is invertible. Then by Lemma 3 of \citet{zamir1998Fisher},
$\mathcal{I}_{z_k} (\bar{Y}_k) = \mathcal{I}_{S_k^+ \ell_k \ell_k^\top z_k} (\bar{Y}_k)
\leq \mathcal{I}_{\ell_k^\top z_k} (\bar{Y}_k)
\leq \mathcal{I}_{z_k} (\bar{Y}_k)$,
%\begin{align*}
%	\mathcal{I}_{z_k} (\bar{Y}_k) = \mathcal{I}_{S_k^+ \ell_k \ell_k^\top z_k} (\bar{Y}_k)
%	\leq \mathcal{I}_{\ell_k^\top z_k} (\bar{Y}_k)
%	\leq \mathcal{I}_{z_k} (\bar{Y}_k),
%\end{align*}
which together with $\ell_k^\top z_k = \ell_k^\top U_k^\top \bar{Y}_{k-1} + \ell_k^\top S_k y_k + \ell_k^\top d_k$
%	\begin{align*}
	%		\ell_k^\top z_k = \ell_k^\top U_k^\top \bar{Y}_{k-1} + \ell_k^\top S_k y_k + \ell_k^\top d_k
	%	\end{align*}
implies
\begin{align*}
	\mathcal{I}_{z_k} (\bar{Y}_k)
	= & \mathcal{I}_{\ell_k^\top z_k} (\bar{Y}_k) 
	= \begin{bmatrix}
		U_k \\ S_k
	\end{bmatrix} 
	\ell_k \left( \ell_k^\top S_k \ell_k \right)^{-1} \ell_k^\top 
	\begin{bmatrix}
		U_k^\top & S_k
	\end{bmatrix} \\
	= & \begin{bmatrix}
		U_k \\ S_k
	\end{bmatrix} 
	S_k^+ 
	\begin{bmatrix}
		U_k^\top & S_k
	\end{bmatrix}
	= \begin{bmatrix}
		U_k S_k^+ U_k^\top & U_k \\ U_k^\top & S_k
	\end{bmatrix}.
\end{align*}
Therefore by \eqref{eq:barS}, $\mathcal{I}_{\bar{Z}_k}(\bar{Y}_k) = \bar{S}_k$. 

Secondly, we will prove $\mathcal{PI}_k \hat{\theta}_k = \bar{H}_k^\top B_k \zeta_k $ by induction. 
Assume that $\mathcal{PI}_{k-1} \hat{\theta}_{k-1} = \bar{H}_{k-1}^\top B_{k-1} \zeta_{k-1} $. Then,
\begin{align*}
	\mathcal{PI}_k \hat{\theta}_k
	= & \mathcal{PI}_k \hat{\theta}_{k-1} +  G_k z_k - G_k \Lambda_k \Psi_k^\top \zeta_k - G_k \Lambda_k G_k^\top \hat{\theta}_{k-1} \\
	= & \bar{H}_{k-1}^\top B_{k-1} \zeta_{k-1} +  G_k z_k - G_k \Lambda_k \Psi_k^\top \zeta_k \\
	= & \bar{H}_{k-1}^\top B_{k-1} \zeta_{k-1} + \bar{H}_{k}^\top \begin{bmatrix}
		B_{k-1} & 0 \\ 0 & I
	\end{bmatrix} \Phi_k z_k \\
	& - \bar{H}_k^\top \Gamma_k \Lambda_k \Psi_k^\top \zeta_k \\
	= & \bar{H}_{k}^\top \begin{bmatrix}
		B_{k-1} & 0 \\ 0 & I
	\end{bmatrix} \zeta_k - \bar{H}_k^\top \Gamma_k \Lambda_k \Psi_k^\top \zeta_k
	= \bar{H}_k^\top B_k \zeta_k.
\end{align*}

Finally, we will prove the theorem. 
Note that by $U_k S_k^+ S_k = U_k$,
\begin{align*}
	\mE \Phi_k z_k 
	= & \mE \begin{bmatrix} U_k S_k^+ \\ I \end{bmatrix} \left( U_k^\top \bar{H}_{k-1} \theta + S_k H_k \theta \right) \\
	= & \begin{bmatrix}
		U_k S_k^+ U_k^\top & U_k \\
		U_k^\top & S_k
	\end{bmatrix} \bar{H}_{k} \theta.
\end{align*}
Then, one can get $\mE \left[ \zeta_k \middle| \theta \right] = \bar{S}_k \bar{H}_{k} \theta$
%\begin{align}\label{eq:mean_zeta}
%	\mE \left[ \zeta_k \middle| \theta \right] = \bar{S}_k \bar{H}_{k} \theta
%\end{align}
by induction. Therefore by Lemma \ref{lemma:PPCR_S}, $\bar{H}_{k}^\top B_k \bar{S}_k \bar{H}_{k} = \mathcal{PI}_k$, which implies the unbiasedness of $ \hat{\theta}_k$. 

Similarly,
% to \eqref{eq:mean_zeta}, 
one can get $\mE \left[ \zeta_k \middle| \bar{Y}_k, \theta \right] = \bar{S}_k \bar{Y}_k $. Define $d_{\Phi,k} = \begin{bmatrix}
	d_{\Phi,k-1} \\ 0
\end{bmatrix} + \Phi_k d_k,\ 
d_{\Phi,0} = \varnothing $.
%\begin{align*}
%	d_{\Phi,k} = \begin{bmatrix}
%		d_{\Phi,k-1} \\ 0
%	\end{bmatrix} + \Phi_k d_k,\quad
%	d_{\Phi,0} = \varnothing. 
%\end{align*}
Then, $\zeta_k = \bar{S}_k \bar{Y}_k + d_{\Phi,k} $. By the independence of $\bar{Y}_k$ and $d_k$, 
\begin{align*}
	& \mE \left[ \left( \zeta_k - \bar{S}_k \bar{H}_k \theta \right) \left( \zeta_k - \bar{S}_k \bar{H}_k \theta \right)^\top \middle| \theta \right] \\
	= & \mE \left[ \left( \bar{S}_k \bar{Y}_k - \bar{S}_k \bar{H}_k \theta \right) \left( \bar{S}_k \bar{Y}_k - \bar{S}_k \bar{H}_k \theta \right)^\top \middle| \theta \right] + \mE d_{\Phi,k} d_{\Phi,k}^\top \\
	= & \bar{S}_k \bar{\Sigma}_{w,k} \bar{S}_k + \mE d_{\Phi,k} d_{\Phi,k}^\top. 
\end{align*}
Since $\mE d_{\Phi,k} d_{\Phi,k}^\top 
= \begin{bmatrix}
	\mE d_{\Phi,k-1} d_{\Phi,k-1}^\top & 0 \\ 0 & 0
\end{bmatrix} + \Phi_k S_k \Phi_k^\top$,
%\begin{align*}
%	\mE d_{\Phi,k} d_{\Phi,k}^\top 
%	= \begin{bmatrix}
%		\mE d_{\Phi,k-1} d_{\Phi,k-1}^\top & 0 \\ 0 & 0
%	\end{bmatrix} + \Phi_k S_k \Phi_k^\top,
%\end{align*}
by \eqref{eq:recur_barS}, one can get $\mE d_{\Phi,k} d_{\Phi,k}^\top = \bar{S}_k$ by induction. Therefore, $\mE \left[ \left( \zeta_k - \bar{S}_k \bar{H}_k \theta \right) \left( \zeta_k - \bar{S}_k \bar{H}_k \theta \right)^\top \middle| \theta \right]
= \bar{S}_k \bar{\Sigma}_{w,k} \bar{S}_k +  \bar{S}_k$.
%\begin{align*}
%	\mE \left[ \left( \zeta_k - \bar{S}_k \bar{H}_k \theta \right) \left( \zeta_k - \bar{S}_k \bar{H}_k \theta \right)^\top \middle| \theta \right]
%	= \bar{S}_k \bar{\Sigma}_{w,k} \bar{S}_k +  \bar{S}_k. 
%\end{align*}
According to $B_k = (I + \bar{S}_k \bar{\Sigma}_{w,k})^{-1} $ and Lemma \ref{lemma:PPCR_S},
\begin{align*}
	& \mE \left[ \left( \hat{\theta}_k - \theta \right) \left( \hat{\theta}_k - \theta \right)^\top \middle| \theta \right] \\
	= & \Sigma_{\text{PPCR},k} \bar{H}_k^\top B_k \left( \bar{S}_k \bar{\Sigma}_{w,k} \bar{S}_k +  \bar{S}_k \right) B_k^\top \bar{H}_k  \Sigma_{\text{PPCR},k} \\
	= & \Sigma_{\text{PPCR},k} \bar{H}_k^\top B_k \bar{S}_k \bar{H}_k  \Sigma_{\text{PPCR},k} \\
	= & \Sigma_{\text{PPCR},k} \mathcal{PI}_k \Sigma_{\text{PPCR},k} 
	= \Sigma_{\text{PPCR},k}. 
\end{align*}
Then, we have the optimality of Algorithm \ref{algo:PP_RLS}. 
\hfill $\qed$

\textbf{Proof of Theorem \ref{thm:rate_attainablility}.}
For notational convenience, in the proof, $\hat{\theta}_{\text{ML}}(\bar{Y}_K) $ is abbreviated as $\hat{\theta}_{\text{ML}}$. 

By Theorem 3.1.1 of \citet{ibragimov1981statistical}, $\hat{\theta}_{\text{ML}} $ converges to $\theta$ almost surely, and $\mE \Absl{\hat{\theta}_{\text{ML}} - \theta}^4 = O(\frac{1}{K^2})$. Additionally, we can use Edgeworth expansion to prove that
\begin{align}\label{eq:ML_2nd}
	\mE\left[ \left(\hat{\theta}_{\text{ML}}- \theta\right)\left(\hat{\theta}_{\text{ML}} - \theta\right)^\top \middle| \theta \right] 
	= \Sigma_{\text{\rm CR},K} + O\left(1/K^2\right). 
\end{align}
The proof of \eqref{eq:ML_2nd} is analogous to \citet{takagi1994second} and is omitted here. Then, \eqref{concl:ML} can be obtained by \eqref{algo:mult}. Therefore, it suffices to prove that $z_K$ in \eqref{algo:mult} satisfies $ \lim_{K\to \infty} \trace(\bar{S}_K) = s_K$ almost surely. 

Due to the differentiability of $l(\cdot)$, it follows that
$\sum_{k=1}^K l^\prime(y_k - H_k \hat{\theta}_{\text{ML}}) H_k^\top = 0$.
%\begin{align*}
%	\sum_{k=1}^K l^\prime(y_k - H_k \hat{\theta}_{\text{ML}}) H_k^\top = 0.
%\end{align*}
Then by the implicit function theorem \citep{zorich2015math},
\begin{align*}
	\frac{\partial \hat{\theta}_{\text{ML}}}{\partial y_k} 
	=&\ \Psi_K^{-1} l^{\prime\prime}(y_k - H_k \hat{\theta}_{\text{ML}}) H_k^\top,\\ 
	\Psi_K =& \sum_{k=1}^K l^{\prime\prime} (y_k - H_k \hat{\theta}_{\text{ML}}) H_k^\top H_k.
\end{align*}

By Lemma 4 of \citet{zamir1998Fisher}, 
\begin{align*}
	\bar{S}_K = \mathcal{I}_{z_K}(\bar{Y}_K)
	= \left[\frac{\partial \hat{\theta}_{\text{ML}}}{\partial \bar{Y}_K} \right]^\top \mathcal{I}_{z_K}(\hat{\theta}_{\text{ML}}) \left[\frac{\partial \hat{\theta}_{\text{ML}}}{\partial \bar{Y}_K} \right]. 
\end{align*}
Therefore,
\begin{align}\label{eq:traceS}
	& \trace(\bar{S}_K) = \trace\left(\mathcal{I}_{z_K}(\hat{\theta}_{\text{ML}}) \left[\frac{\partial \hat{\theta}_{\text{ML}}}{\partial \bar{Y}_K} \right] \left[\frac{\partial \hat{\theta}_{\text{ML}}}{\partial \bar{Y}_K} \right]^\top \right) \nonumber\\
	= & \trace\left(\frac{\mathcal{I}_{z_K}(\hat{\theta}_{\text{ML}})}{K} \cdot K \sum_{k=1}^K \left[\frac{\partial \hat{\theta}_{\text{ML}}}{\partial y_k} \right] \left[\frac{\partial \hat{\theta}_{\text{ML}}}{\partial y_k} \right]^\top \right). 
\end{align}

By Assumption \ref{assum:mult} i) and \eqref{algo:mult}, 
\begin{align}\label{eq:lim_Fisher/K}
	\lim_{K\to\infty} \frac{\mathcal{I}_{z_K}(\hat{\theta}_{\text{ML}})}{s_K K} 
	= & \lim_{K\to\infty} \frac{(\mE l^{\prime\prime}(w))^2}{K n \mE (l^{\prime\prime}(w))^2} \sum_{k=1}^K H_k^\top H_k \nonumber\\
	= & \frac{(\mE l^{\prime\prime}(w))^2 \Sigma_H}{n \mE (l^{\prime\prime}(w))^2},\ \as
\end{align}
Additionally, by the law of large numbers \citep{shao2003mathematical} and the almost sure convergence of $\hat{\theta}_{\text{ML}}$,
\begin{align*}
	& \lim_{K\to \infty} \frac{\Psi_K}{K}
	= \lim_{K\to \infty} \frac{\sum_{k=1}^K l^{\prime\prime} (y_k - H_k \hat{\theta}_{\text{ML}}) H_k^\top H_k}{K} \\
	= & \lim_{K\to \infty} \frac{\sum_{k=1}^K l^{\prime\prime} (y_k - H_k \theta) H_k^\top H_k}{K} \\
	= & \lim_{K\to \infty} \frac{\sum_{k=1}^K \left(l^{\prime\prime}(w_k) - \mE l^{\prime\prime}  (w_k)\right) H_k^\top H_k}{K} \\
	& + \lim_{K\to \infty} \frac{\sum_{k=1}^K \mE l^{\prime\prime}  (w_k) H_k^\top H_k}{K}
	= \mE l^{\prime\prime}  (w) \Sigma_H,\ \as 
\end{align*} 
Similarly, 
\begin{align*}
	\lim_{K\to \infty} \frac{\sum_{k=1}^K (l^{\prime\prime} (y_k - H_k \hat{\theta}_{\text{ML}}))^2 H_k^\top H_k}{K} = \mE (l^{\prime\prime}  (w))^2 \Sigma_H,\ \as 
\end{align*}
Therefore, we have
\begin{align*}
	& \lim_{K\to \infty} K \sum_{k=1}^K \left[\frac{\partial \hat{\theta}_{\text{ML}}}{\partial y_k} \right] \left[\frac{\partial \hat{\theta}_{\text{ML}}}{\partial y_k} \right]^\top \\
	= & \lim_{K\to \infty} \left( \frac{\Psi_K}{K} \right)^{-1} \frac{\sum_{k=1}^K (l^{\prime\prime} (y_k - H_k \hat{\theta}_{\text{ML}}))^2 H_k^\top H_k}{K} \left( \frac{\Psi_K}{K} \right)^{-1} \\
	= & \frac{\mE (l^{\prime\prime}  (w))^2}{(\mE l^{\prime\prime}  (w))^2} \Sigma_H^{-1},\ \as, 
\end{align*}
which together with \eqref{eq:traceS} and \eqref{eq:lim_Fisher/K} implies
\vspace{-0.2em}
\begin{align*}
	\lim\limits_{K\to\infty} \frac{\trace(\bar{S}_K)}{s_K} = \frac{1}{n} \trace(I_n) = 1,\ \as 
\end{align*}
Note that $d_K$ is independent of $\bar{Y}_K$. Then, the theorem can be proved by \eqref{eq:ML_2nd}. 
\hfill $\qed$

\end{document}